\documentstyle[12pt,twoside]{article}
\def\be{\begin{equation}}
\def\ee{\end{equation}}
\def\bea{\begin{eqnarray}}
\def\eea{\end{eqnarray}}
\def\nn{\nonumber}
\def\half{\frac{1}{2}}
\def\a{\alpha}
\def\b{\beta}

\def\d{\partial}
\def\de{\delta}
\def\e{\epsilon}           
\def\g{\gamma}

\def\l{\lambda}
\def\m{\mu}
\def\n{\nu}
\def\o{\omega}
\def\r{\rho}                      
\def\s{\sigma}                    
\def\t{\tau}

\def\w{\o}

\def\G{\Gamma}

\def\O{\Omega}

\def\Q{\Theta}
\def\S{\Sigma}

\def\widebar{\overline}
\def\ul{\underline}
\def\ol{\widebar}

\def\calL{{\cal L}}

\def\diag{{\rm diag}}

\def\psibar{\widebar{\psi}}
\def\phibar{\widebar{\phi}}

\def\neta{\eta}

\def\R{{\rm R}}

\def\divrt{\frac{1}{\sqrt{2}}}

\def\one{{\bf 1}}

\def\calL{{\cal L}}
\newcommand{\wt}[1]{\widetilde{#1}}
\newcommand{\what}[1]{\widehat{#1}}
\def\dehat{\what{\de}}

\begin{document}
\newcommand{\rf}[1]{~(\ref{#1})}

\thispagestyle{empty}
\begin{flushright}
{\sc ITP-SB}-97-4

{\sc HUB-EP} 97/7
\end{flushright}
\vspace{.1cm}
\setcounter{footnote}{0}
\begin{center}
{\Large\bf{
On a possible new $R^2$ theory of
supergravity.~\footnote{This research was 
supported in part by NSF 
grant no PHY9309888 and the Swiss National Science Foundation.}
}}\\[7mm]

\bf{Tobias 
Hurth\footnote{e-mail: hurth@insti.physics.sunysb.edu},
 Peter van 
Nieuwenhuizen\footnote{e-mail: vannieu@insti.physics.sunysb.edu},
Andrew Waldron\footnote{e-mail:
wally@insti.physics.sunysb.edu}}\\[1mm]
{\it Institute for Theoretical Physics,\\
State University of New York at Stony Brook,\\
Stony Brook, NY 11794-3840, USA.}\\[2mm]
\bf{and}\\[2mm]
\bf{Christian
Preitschopf\footnote{e-mail:
preitsch@qft2.physik.hu-berlin.de}}\\[1mm]
{\it Institut f\"ur Physik,\\
Humboldt-Universit\"at zu Berlin,\\
Invalidenstr. 110, D-10115 Berlin, Germany}\\[4mm]

{\sc Abstract}\\[1mm]
\end{center}

We consider a new MacDowell--Mansouri $R^2$--type of
supergravity theory, an extension of conformal supergravity,
based on the superalgebra $Osp(1|8)$.
Invariance under local symmetries with negative Weyl weight is
achieved by imposing chirality-duality and double-duality
constraints on curvatures, along with the usual constraint of
vanishing supertorsion. An
analysis of the remaining gauge
symmetries shows that those with vanishing Weyl weight are
invariances of the action at the linearized level. For
the symmetries with positive Weyl weight we find that 
invariance of the action 
would require further modifications of the transformation rules.
This conclusion is supported by a kinematical analysis 
of the closure of
the gauge algebra.

\newpage

\section{Introduction and Review.}
\label{sec1}

This article presents a study of a possible extension of 
conformal supergravity in four dimensions. The motivation of our
work is the growing belief, due to string theory, that new
supergravities in 10, 11 and or 12 dimensions may
exist~\cite{azcar}-\cite{11and12}.
It is
known~\cite{Haag} 
that the largest symmetry of the $S$--matrix in
asymptotically flat spacetime, is the superconformal algebra,
therefore, if the model were to exist as a
quantum theory, the extra symmetries of the action would not
manifest themselves as symmetries of the $S$--matrix.

In this article we try to construct a classical action,
quadratic in curvatures, which is invariant under all local
symmetries of the superalgebra $Osp(1|8)$. The algebra is
given in\rf{qq6} and\rf{quirk}-(\ref{quark}), the action is given 
in\rf{S} and the constraints on curvatures are given
in\rf{mc1}-(\ref{mc5}). 
We use an approach based on the ``gauging'' of spacetime algebras
by imposing constraints on curvatures which was published in
this journal~\cite{Kaku1}.
We have obtained invariance under most
local symmetries, but further modifications of the Yang--Mills
rules for gauge theories are found to be necessary, see the last
section. 
A hint in this direction is the observation that if one uses the
results of~\cite{PvN6} that in ordinary conformal supergravity in
$3+1$ dimensions the number of bosonic and fermionic states
match, and notices that the fermionic spectrum of $Osp(1|8)$
is the same as in the superconformal case, it follows that there
should not be any further bosonic states if there should be
equal numbers of bosonic and fermionic states. 
We now begin with an introduction to the
``gauging of spacetime algebras''; some of this material is new,
while the rest is based on~\cite{MacDowell}-\cite{PvN10}.

When supergravity was discovered, it was
constructed as the gauge theory of supersymmetry,
and contained only two physical fields: the spin 2 gravitational
vierbein field $e_\m{}^m$ and its superpartner, the spin $3/2$
supersymmetry gauge field $\psi_\m{}^\a$~\cite{PvN1}.
The spin connection $\o_\m{}^{mn}$ (the gauge field for Lorentz
symmetry) was expressed in terms of $e_\m{}^m$, as usual
in general relativity. The action consisted of three
parts: the Einstein-Hilbert action (with
vierbein dependent $\w(e)_\m{}^{mn}$),
the Rarita-Schwinger action for the spin $3/2$ gauge
field $\psi_\m{}^\a$ (with a Lorentz covariant derivative
containing $\o(e)_\m{}^{mn}$), and complicated four fermion
terms. In a reformulation, the theory was written with an
independent spin connection $\o_\m{}^{mn}$, and in this
formulation there were no four-fermion terms~\cite{PvN2}. 
However upon solving the algebraic field equation for
$\o_\m{}^{mn}$ one finds
$\w_\m{}^{mn}$ $=$ $\w(e)_\m{}^{mn}+\psibar\psi$ terms
and substituting this solution back into the action, the
formulation with $\o_\m{}^{mn}(e)$ and four-fermion terms
reappeared. As already anticipated in~\cite{PvN3}, the spin 2 and
spin 3/2 curvatures which appear in the action 
are Yang--Mills curvatures 
for the super Poincar\'{e} algebra, but not
anticipated in~\cite{PvN3}, is the fact that ``gauging'' of the
super Poincar\'{e} algebra is not enough: 
in order to obtain invariance
under local supersymmetry one must either use an independent
spin connection whose transformation rule is {\it not}
the same as obtained by gauging the super Poincar\'{e}
algebra, {\it or} one should use the expression
$\w(e,\psi)_\m{}^{mn}$ which results from solving the
$\o$ field equation.
Only gradually realized after~\cite{PvN1} and~\cite{PvN2}
is the fact that this result for $\o(e,\psi)_\m{}^{mn}$ can also
be obtained by imposing the constraint that
\be
R(P)_{\m\n}{}^m=0,\ee
where $R(P)_{\m\n}{}^m$ is the Yang--Mills curvature for the
translation generator in the super Poincar\'{e} algebra.
Thus the dynamical origin of the compositeness of
$\o(e,\psi)_\m{}^{mn}$ was replaced by a kinematical origin. The
relation between first-order formalism (with independent
$\w_\m{}^{mn}$) and second-order formalism (with
$\o(e,\psi)_\m{}^{mn}$) becomes very clear by writing the
variation of the action under supersymmetry as
\be
\de S\sim
\int\e^{\m\n\r\s}\e_{mnrs}R(P)_{\m\n}{}^m\left[\de
\o_\r{}^{nr}-\O(e,\psi)_\r{}^{nr}\right]e_\s{}^s.
\ee
In this expression the variation of $\de e_\m{}^m$ and
$\de\psi_\m{}^\a$ has already been performed, giving rise to
$\O_\r{}^{nr}$, which is a complicated function of $e_\m{}^m$ and
$\psi_\m{}^\a$. Invariance under local supersymmetry
($\de S=0$) can be obtained in two
(and not more) ways: either by setting $R(P)_{\m\n}{}^m=0$
(second-order formalism), or by setting
$\de\o_\r{}^{nr}=\O_\r{}^{nr}$ (first-order formalism).
A very useful observation, sometimes called 1.5 order formalism,
states that since in second-order formalism
$\o(e,\psi)$ satisfies its own field equation,
it is not necessary to vary the composite object $\o(e,\psi)$
(because if one were to vary it using the chain rule, the result
would anyhow be multiplied by the $\o$ field equation which
vanishes {\it identically} in second-order formalism). 

Another reformulation of this simplest
supergravity theory ($N=1$, or simple supergravity) was
discovered by MacDowell and Mansouri~\cite{MacDowell}. Instead of
the super Poincar\'{e} algebra, they considered the super-anti
de Sitter algebra, with the same generators $P_m$ (for
translations), $Q_\a$ (for supersymmetry) and $M_{mn}$ (for
Lorentz symmetry), but with different curvatures:
$R(P)_{\m\n}{}^m$ is the same, but $R(M)_{\m\n}{}^{mn}$ has 
two extra terms
$-\l^2e_\m^{[m}e_\n{}^{n]}+\l\psibar_\m\g^{[m}\g^{n]}\psi_\n$
and $R(Q)_{\m\n}{}^\a$ has an extra term
$\begin{array}{c}\frac{1}{2}\end{array}\l(\g_\m\psi_\n{}^\a-\g_\n
\psi_\m{}^\a)$.
The constant $\l$ is the inverse of the radius of the anti de
Sitter space. They proposed the following action
\bea
S&=&\int d^4x\e^{\m\n\r\s}[R(M)_{\m\n}{}^{mn}R(M)_{\r\s}{}^{rs}
\e_{mnrs}\nn\\
&&\hspace{.5cm}+8\l R(Q)_{\m\n}^\a\g_{\a\b}^5R(Q)_{\r\s}{}^\b].
\eea
Expanding
$R(M)_{\m\n}{}^{mn}=R(M)^{\rm L}_{\m\n}{}^{mn}+\l$--dependent
terms, where $R(M)^{\rm L}_{\m\n}{}^{mn}$ is the Lorentz
curvature, they found that the leading term in the first square
is the Gauss--Bonnet topological invariant
($R(M)_{\rm L}^{mn}\wedge R(M)_{\rm L}^{rs}\e_{mnrs}$). The 
cross term between
$R(M)^{\rm L}$ and $\l\psibar\psi$ cancels a similar term due to
partially integrating one of the two covariant derivatives
$D(\o)$ in the second square of curvatures. One is then left 
with: the Einstein-Hilbert action ($R(M)^{\rm L}$ 
contracted with the
$\l^2ee$ term), the Rarita-Schwinger action ($R(Q)$
contracted with the term $\l\g\psi$) and a supercosmological
constant (the square of $\l^2ee+\l\psibar\psi$ plus the square
of $\l\g\psi$). By first dropping the topological term, then
dividing by $\l$ and finally
taking the limit $\l\rightarrow0$, the ordinary $N=1$
supergravity theory was reobtained.

This approach was extended to $N=2$ supergravity whose physical
fields are all gauge fields ($e_\m{}^m$, $\psi_\m{}^{\a a}$
with $a=1,2$, and $A_\m$)~\cite{PvN5}.
However, for $N\geq3$ supergravity no similar results could
be obtained since these theories contain also scalar fields.

Subsequently, Kaku, Townsend and van Nieuwenhuizen  
studied supersymmetric conformal gauge theories, instead of
supersymmetric Poincar\'{e} gauge
theories~\cite{Kaku,Kaku1}. 
Since the conformal
group contains dilations $D$, the action should not 
contain dimensionful parameters, and since the dimension of the
Lagrangian density is 4 while
curvatures (corresponding to the zero dilaton weight gauge
fields) have dimension 2, it
was natural to construct an $R^2$ action for conformal
supergravity along the lines of MacDowell and Mansouri. There are
now $15+9=24$ generators, which can be written in terms of
increasing scale weight as
\be
\begin{array}{ccccc}P_m(e_\m{}^m)&Q_\a(\psi_\m{}^\a)&M_{mn}(\o_\m
{}^{mn})&S_\a(\phi_\m{}^\a)&K_m(f_\m{}^m)\\
&&D(b_\m)&&\\&&A(a_\m)&&
\end{array}
\ee
The corresponding gauge fields are given in parentheses.
The $S_\a$ are conformal supersymmetry generators, the $K_m$
denote conformal boosts while the bosonic generator $A$ for
chiral transformations is also needed to extend the ordinary
conformal algebra $SU(2,2)$ to the superconformal algebra
$SU(2,2|1)$. As action they took the most general dilaton-weight
zero, Lorentz and Einstein scalar
\bea
&\int
d^4x[aR(P)R(K)+bR(Q)\g_5R(S)+cR(M)\wt{R}(M)+dR(D)R(A)+&\nn\\
&\a eg^{\m\r}g^{\n\s}R(A)_{\m\n}R(A)_{\r\s}].&
\eea
(where $a$, $b$, $c$, $d$ and $\a$ are constants and $\wt{R}(M)$
denotes
$\begin{array}{r}\frac{1}{2}\end{array}\e^{mnrs}$ $R(M)^{r's'}$
$\eta_{rr'}\eta_{ss'}$) 
All terms were affine (i.e., indices were contracted with
constant Lorentz-invariant tensors) except the last term.
The last term was added since by counting {\it states}
one finds~\cite{PvN6} that the chiral gauge field $A_\m$ should
be physical, and the term $R(D)R(A)$ seemed not to lead to a
propagator for $A_\m$.

By construction the action was invariant under $M$, $D$, $A$
symmetries and under general coordinate transformations.
Requiring invariance under the Yang--Mills symmetries
corresponding to $S$ and $K$ led to the following 
information~\cite{Kaku1} 
\begin{enumerate}
\item constraints on the curvatures were needed:
$R(P)_{\m\n}{}^m=0$ and the chirality-duality constraint
$R(Q)_{\m\n}+\!\!\begin{array}{r}\frac{1}{2}\end{array}
\g^5e\e_{\m\n\r\s}R(Q)^{\r\s}=0$.

\item all constants were fixed.
\end{enumerate}
With hindsight, this was only one possible solution. When the
analysis of invariance under $S$ and $K$ was performed, it was
not yet known which further constraints on curvatures would be
found. In particular, $R(A)_{\m\n}$ and $R(D)_{\m\n}$ were still
considered as independent objects, and this fixed both $\a$ and
$d$. Later, a constraint $R(A)={}_*R(D)$ was found (where
${}_*R(D)_{\m\n}=\!\!\begin{array}{r}\frac{1}{2}\end{array}e
\e_{\m\n\r\s}R(D)_{\r'\s'}
g^{\r\r'}g^{\s\s'}$), and using this in 
the $K$ and $S$ variations, a one-parameter solution is
found. Substituting $R(A)_{\m\n}={}_*R(D)_{\m\n}$ 
in the action, this freedom
is clearly cancelled.

Requiring invariance under $Q$ (ordinary local supersymmetry)
was a complicated task, but finally achieved. A further
constraint $\g^{\m\n}R(Q)_{\m\n}=0$ was found to be necessary,
and the 1.5 order formalism for the conformal gauge field
$f_\m{}^m$ corresponding to $K_m$ was used: $f_\m{}^m$ had an
algebraic field equation, and upon solving it one found
$f_\m{}^m$ as a complicated 
composite field $f_\m{}^m(e,\psi,\phi,\o,b,A)$ whose variation
was not needed for reasons explained above.
The two constraints for $R(Q)_{\m\n}$ could be written together
in the simple form
\be
\g^\m R(Q)_{\m\n}=0
\ee 
and these 16 constraints could be solved, yielding the conformal
supersymmetry gauge field $\phi_\m{}^\a$ as a composite object. 
As in
ordinary supergravity (and gravity), the 24 constraints
$R(P)_{\m\n}{}^m=0$ yielded $\o_\m{}^{mn}$ as a composite
object. This left only the fields $e_\m{}^m$, $\psi_\m{}^\a$,
$A_\m$ and $b_\m$ as independent fields. However, as a result of
$K$--gauge invariance, the dilaton gauge field $b_\m$ dropped
from the action, and one ended up with only $e_\m{}^m$,
$\psi_\m{}^\a$ and $a_\m$. (In later developments, Poincar\'{e}
supergravity was obtained from conformal supergravity by
coupling the latter to a WZ multiplet. The axial gauge field
$a_\m$ and the WZ auxiliary fields $F_{WZ}$ and $G_{WZ}$ became
then the minimal auxiliary fields $S$, $P$, $A_\m$ of $N=1$
supergravity, while $D$, $A$ and $S$ symmetry was fixed by
choosing the gauge $A_{WZ}=1$ and $B_{WZ}=\chi_{WZ}=0$.
Since $K$-symmetry did not act on $e_\m{}^m$, $\psi_\m{}^\a$
or $A_\m$ it did not need to be fixed~\cite{KTD}). 

A simplification was obtained when it was realized~\cite{PvN7}
 that
the $f_\m{}^m$ 
field equation could also be written as a constraint on the
Ricci tensor $e_m{}^\m R(M)_{\m\n}{}^{mn}$, just as in ordinary
supergravity where 
the $\o$ field equation could be written as the
constraint $R(P)_{\m\n}{}^m=0$.
At this point there were 3 constraints, $R(P)_{\m\n}{}^m=0$,
$\g^\m R(Q)_{\m\n}=0$ and $e_m{}^\m R(M)_{\m\n}{}^{mn}+\cdots=0$
(where $\cdots$ contain supercovariantizations, see 
eqn.\rf{sfeqn}); 
these constraints are all field equations in Poincar\'{e}
supergravity.

A final simplification was obtained in~\cite{PvN}, where only the
pure affine part of the action was considered (so with $\a=0$).
It was found that the constraint on the Ricci tensor implied
another duality constraint
\be
R(A)_{\m\n}=\frac{1}{2}e\e_{\m\n\r\s}R^{\r\s}(D),
\ee
where $e=\det e_\m{}^m$,
so that the term $R(A)\wedge R(D)$ did yield, after all, the
Maxwell action for the chiral field $a_\m$.
However, for the affine action, the constraint on the Ricci
tensor is no longer a field equation.

Up till now we have reviewed and commented on the dynamical
approach (requiring invariance of the action), because this is
the approach we shall mainly follow below. However, in the
original work~\cite{Kaku,Kaku1} 
and especially in~\cite{PvN}, a kinematical
approach was followed which directly yielded all constraints. We
briefly review it here since we shall use it in the last section
of this paper. 

The basic idea~\cite{Kaku1} is that 
when the (anti)commutator of two
Yang--Mills symmetries ($P$--gauge 
Yang--Mills symmetries excluded)
produces a $P$--gauge Yang--Mills 
symmetry, constraints on one or more
curvatures are necessary.  These 
constraints should be such that as a
result, one or more fields become 
dependent fields and their
transformation rules (obtained by 
applying the chain rule) should
deviate from the Yang--Mills 
transformation rule (which uses only the
structure constants of the superalgebra) 
in such a way that one finds
on the right hand side a general coordinate 
transformation instead of
a $P$--gauge Yang--Mills transformation. 
Starting with the gauge fields
with highest dilaton weight ($e_\m{}^m$), 
and then moving towards
fields with decreasing dilaton weight, and 
requiring that the algebra
now closes onto general coordinate rather than $P$--gauge
transformations, one finds, {\it in an unambiguous and methodical
way}, all constraints on curvatures.

For the conformal superalgebra, one only needs to consider the
anticommutator of two 
$Q$'s since $\{Q,Q\}\sim P$, and one acts
successively on $e_\m{}^m$, 
$\psi_\m{}^\a$, $b_\m$ and $a_\m$.  This
produces the constraints 
$R(P)_{\m\n}{}^m=0$, $\g^\m R(Q)_{\m\n}=0$
and the constraint on the Ricci tensor, respectively, 
and solving
these constraints, $\o_\m{}^{mn}$, $\phi_\m{}^\a$ and $f_\m{}^m$
become dependent fields.

Work on conformal supergravity continued in 3
dimensions~\cite{PvN10}
where it turned out to yield Chern-Simons theory, but this will
not concern us here.

Recently, there have been speculations that further supergravity 
theories might exist in $d=11$ and $d=12$
dimensions.
These ideas are based on the fact that p-branes couple naturally
to (p+1)-form gauge potentials via the 
currents~\cite{azcar,pktrev}:
\be
\begin{array}[t]{rcl}
J^{\m_1 \ldots \m_{p+1}}(x) &= \frac{1}{\sqrt{g}} & 
\!\!\!\!\int d\t \int d^p \s
\ \de^d (x - X(\t,\s) ) \\ &&
\ \e^{i_1 \ldots i_{p+1}} \d_{i_1} X^{\m_1}(\t, \s) \ldots  
 \d_{i_{(p+1)}} X^{\m_{(p+1)}}(\t, \s) \ .
\end{array}
\ee
These currents are conserved:
\be
\d_\n ( \sqrt{g} J^{\n \m_1 \ldots \m_{p}}(x) ) \ = \ 0
\ee
and give rise to tensor charges
\be
Z^{\m_1 \ldots \m_{p}} \ = \ \int d^{d-1} x J^{0 \m_1 
\ldots \m_{p}}(x) \ .
\ee
They appear then in (maximally) extended supersymmetry algebras
as follows~\cite{pktdem}: for the case of IIA supersymmetry 
in d=(1,9) one has
\bea
\hspace{-.3cm} \{ Q_{a} , Q_{b}\} &=&
\begin{array}[t]{r}
\G_{ab}^\m P_\m + \G_{ab}^\m Z_\m + 
\G_{ab}^{\m_1 \ldots \m_5} Z^+_{\m_1 \ldots \m_5} 
\end{array}\\
\hspace{-.3cm} \{ Q^{a} , Q^{b}\} &=&
\begin{array}[t]{r}
\G^{\m ab} P_\m - \G^{\m ab} Z_\m + 
\G^{\m_1 \ldots \m_5 ab} Z^-_{\m_1 \ldots \m_5} 
\end{array}\\
\hspace{-.3cm} \{ Q_{a} , Q^{b}\} &=&
\begin{array}[t]{r}
\de_a^b Z + \G^{\m\n}{}_a{}^b Z_{\m\n} + 
\G^{\m_1 \ldots \m_4}{}_a{}^b Z_{\m_1 \ldots \m_4} 
\end{array}
\label{twoa}
\eea
where we have 16-dimensional Majorana-Weyl spinors $Q_a$ and
$Q^a$ of opposite chirality.  
This algebra may be given a (1,10)-d 
interpretation in terms of real
32-component spinors:
\bea
\hspace{-.3cm} \{ Q_{\a} , Q_{\b}\} &=&\!\!
\begin{array}[t]{r}
\G_{\a\b}^M P_M + \G_{\a\b}^{M N} Z_{MN} + 
\G_{\a\b}^{M_1 \ldots M_5} Z_{M_1 \ldots M_5}\ , 
\end{array}
\label{elevenalg}
\eea
or even a (2,10)-d one in terms of 32-component Majorana-Weyl
spinors:
\bea
\hspace{-.3cm} \{ Q_{\a} , Q_{\b}\} &=&\!\!
\begin{array}[t]{r}
\G_{\a\b}^{M N} M_{MN} + 
\G_{\a\b}^{M_1 \ldots M_6} Z_{M_1 \ldots M_6}\ . 
\end{array}
\label{twelvealg}
\eea
The type IIB algebra in d=(1,9) reads
\bea 
\hspace{-.3cm} \{ Q_{ai} , Q_{bj}\} &=&\!\!\!\!\!\!\!\!\!
\begin{array}[t]{rl}
& \begin{array}[t]{l}
\G_{ab}^\m \Sigma^J_{ij}  Z_{J\m}
+
\G_{ab}^{\m_1 \m_2 \m_3} \e_{ij} Z_{\m_1 \m_2 \m_3}
+
\G^{\m_1 \ldots \m_5}_{ab} \Sigma^J_{ij} 
Z_{J\m_1 \ldots \m_5},
\end{array}
\end{array}
\label{twob}
\eea
where we use the 
conventions $\S^J_{(ij)} = \e_{il} \S^{Jl}{}_j$, 
$\S_0 = -i \s^2, \S_1 = -\s^1$ and $\S_2 = \s^3$. 
These matrices satisfiy
$\S_I \S_J = 
\eta_{IJ} + 
\e_{IJ}{}^K \S_K = \eta_{IJ} +\e_{IJL}\eta^{LK} \S_K$,
with $\eta_{IJ} = (-++)$, $\e_{012}=1$,  
and hence generate $SL(2,{\bf R})$. 
In all cases the $Z$-charges fit the respective brane-scan, and
all cases form some decomposition of the $Q$-$Q$ part of 
$OSp(1|32)$.
It seems plausible that 
the rest of
the algebra completes (possibly some contraction of) the 
extended
(1,9)-d superconformal algebra of 
van Holten and van Proeyen~\cite{vanPr}, namely
\bea
\hspace{-.3cm} \{ Q_{\a} , Q_{\b}\} &=&
\begin{array}[t]{r}
- \frac{1}{128} \G_{\a\b}^{M N} J_{MN} - \frac{1}{128 \cdot 6!}  
\G_{\a\b}^{M_1 \ldots M_6} J_{M_1 \ldots M_6}\ 
\end{array} \nn \\
\hspace{-.3cm} [ J_{MN} , Q_{\a}] &=&
\begin{array}[t]{r}
- (\G_{MN})_{\a}{}^{\b} Q_\b
\end{array} \nn \\
\hspace{-.3cm} [ J_{M_1 \ldots M_6} , Q_{\a}] &=&
\begin{array}[t]{r}
- (\G_{M_1 \ldots M_6})_{\a}{}^{\b} Q_\b
\end{array} \nn \\
\hspace{-.3cm} [ J_{MN} , J^{KL}] &=&
\begin{array}[t]{r}
8\ \de^{[K}_{[N} \ J_{M]}{}^{L]}
\end{array} \nn \\
\hspace{-.3cm} [ J_{MN} , J^{M_1 \ldots M_6}] &=&
\begin{array}[t]{r}
24\ \de^{[M_1}_{[N} \ J_{M]}^{\ \ M_2 \ldots M_6]}
\end{array} \nn \\
\hspace{-.3cm} [ J_{N_1 \ldots N_6} , J^{M_1 \ldots M_6}] &=&
\begin{array}[t]{r}
\begin{array}[t]{l}
-12 \cdot 6!\  \de^{[M_1 \ldots M_5}_{[N_1 \ldots N_5} \ 
J_{N_6]}^{\ \ M_6]} \vspace{0.1cm} \\ \vspace{0.2cm}
+ 12 \ 
\e_{N_1 \ldots N_6}{}^{[M_1 \ldots M_5|R|} \ 
J_R{}^{M_6]} \ .
\end{array}
\end{array} 
\label{twelvecomplete}
\eea
In the (1,10)-d context this algebra was studied by D'Auria and
Fr\'{e}~\cite{Fre}.  We will try to take some first steps towards
constructing a supergravity theory based on that algebra.

The signature of the vector space that appears in the  
above algebra
is $(2,10)$. This provides another hint of a connection to 
string-theoretic
ideas, as Vafa's~\cite{vafaevi} argument shows: 
$Sl(2,Z)$-duality
of type IIB strings may be explained via D-strings. 
The zero-modes of
the open strings stretched between such D-strings 
determine the worldsheet
fields of the latter. We have
\bea
\Psi^\m_{-1/2} |k> & \m=0,1 & \hbox{2-d gauge fields}\\
\Psi^m_{-1/2} |k> & m=2,\cdots,9 & \hbox{transverse 
fluctuations,}
\label{openmodes}
\eea 
and hence we find on the D-string an extra U(1) gauge field. In
d=2 this is nondynamical, of course, but it leaves, 
after gauge fixing,
a pair of ghosts $B,C$ with central charge $c=-2$. The critical
dimension is hence raised by two, and the no-ghost
theorem~\cite{noghost,nogbrst}, which states that the BRST
cohomology effectively eliminates those extra dimensions, 
forces us to
assume the existence of a nullvector in the extra 
dimensions, and that
means they must have signature (1,1).

Taking the idea of strings moving in a 12-dimensional
target space more seriously, we are immediately led to the puzzle
of why strings oscillate in only 10 of these dimensions, but
never in the extra 2. If one has conformal symmetry in mind, 
there
is a natural answer: the 12 dimensions are those in which the
conformal group is linearly realized, but only a 10-dimensional
null hypersurface in real projective classes of these 
coordinates is 
physical. The extra two dimensions ``don't really exist''. 
The idea that strings might have some sort of target space 
conformal symmetry is not new~\cite{confstring}, but as of now no model
exists that can be convincingly linked to the string theories known 
today. 

The idea of supergravities beyond d=11 was already explored 
fifteen
years ago, but no {\it conventional} supergravity theory was
found~\cite{unconventional}, even though in $d=(10,2)$ dimensions
Majorana-Weyl spinors exist, and dimensional reduction 
to $d=(10,1)$
would therefore lead to a $N=1$ supergravity theory. 
One of us has
been pursuing this problem further over the years, 
also studying a
possible gravitational Chern-Simons supergravity theory in 
$d=11$. In
fact, we might speculate that an unconventional (topological?)
supergravity theory in $d=(10,2)$ might lead to a new $d=11$
supergravity theory (Chern-Simons ?) in which the usual 
problems with
Kaluza-Klein compactification might go away.  However, 
if new theories
do exist in $d=12$ and/or $d=11$, there might perhaps 
also be new
theories in $d=4$, and conversely, finding new theories 
in $d=4$ might
show the way to new theories in $d=11$ and/or $d=12$.



With this motivation in mind, it is natural to ask for a modification
of the superconformal algebra in $d=4$, and to try to construct a
corresponding gauge theory.  A natural candidate is the
four-dimensional version of~\rf{twelvecomplete}, namely $Osp(1|8)$. It
resembles $SU(2,2|1)$ in that the two spinors $Q_\a$ and $S_\a$ form
an 8-real-component spinor in the fundamental representation, but now
of course of $Sp(8,{\bf R})$, which contains as a maximal subalgebra
$U(2,2)=SU(2,2)\times U(1)$, where $SU(2,2)$ is the conformal algebra
and $U(1)$ corresponds to the chiral generator $A$. In fact, since
the bosonic subalgebra of $Osp(1|8)$ is simple, in contrast to the
superconformal case, one would be inclined to believe
that the geometric approach ought to work even better for
$Osp(1|8)$.   

This is certainly true for $Osp(1|4)$: since $Sp(4)$ is isomorphic to
the anti de Sitter algebra $SO(2,3)$, we can view it as the conformal
algebra in $d=(1,2)$. (Recall that $SU(2,2)$ is isomorphic to
$SO(2,4)$).  Thus the conformal algebra in $d=(1,2)$ corresponds to an
ordinary anti de Sitter algebra in $d=(1,3)$, and both algebras can be
supersymmetrized, and lead to gauge theories. In a similiar manner,
one might expect that if an extended superconformal gauge theory for
$Osp(1|8)$ can be found, it will corrrespond to a non-conformal
supergravity theory in 5-dimensional Minkowski spacetime. Jumping
ahead: perhaps $Osp(1|32)$ yields an extended superconformal theory in
10-dimensional Minkowski spacetime, and perhaps this would lead to a
new theory in $d=11$~\cite{cschile}. 

Using the $R^2$-method, we will now start constructing a field theory
based on $Osp(1|8)$ in $d=(1,3)$ spacetime. The harder problem of
$Osp(1|32)$ we temporarily set aside.
In section~\ref{subnuclear} 
we present a real 
representation 
for $Osp(1|8)$ which at once identifies the gauge
fields, curvatures
and the Yang--Mills transformation rules. 
In section~\ref{nuclear} 
we write down the most general affine $R^2$ action 
invariant under all symmetries with positive 
dilaton weight. (We call them ``sure symmetries''. 
They are generalizations of
 $K_\a$ and $S_\a$.) By construction this 
action is also Lorentz, dilation
and chiral invariant (``automatic symmetries''). 
This leaves us with the 
symmetries corresponding to two new scale-zero generators
($V_m$ and $Z_m$), ordinary supersymmetry ($Q_\a$), 
and the local symmetries 
corresponding to two scale-2 generators: $P_m$ and $E_{mn}$.
As before, we exchange $P_m$ Yang--Mills symmetry with 
general coordinate
invariance so that we are left with the problem 
of obtaining local 
$V_m$, $Z_m$, $Q_\a$ and $E_{mn}$ symmetry of the 
action (we call these ``unsure'' symmetries). 
As in the case of conformal supergravity, we anticipate
that we will need further constraints, and with this 
possibility in mind, 
in sections~\ref{Lorentz} 
and~\ref{Myrtle} we construct a list of the maximal set of 
constraints on curvatures which can be solved by expressing one
(or more) field(s) in terms of others. 
It is not obvious at this point that all these constraints 
must actually be imposed, but it is clear that only constraints 
from this set can play 
a r\^{o}le. In sections~\ref{Linear} and~\ref{blueline}, 
we make an analysis of all unsure
symmetries at leading order.
We find that $V_m$ and $Z_m$
invariance in fact requires the maximal 
set of solvable constraints
at the linearized level and we do find invariance under these
symmetries at the linearized level.
In section~\ref{concl}, we give our conclusions, and make a 
purely kinematical 
analysis of $Q$ and $E_{mn}$ symmetry
 by studying the conditions under
which $P_m$ gauge transformations turn into general 
coordinate transformations.

\section{The Algebra and Derivation of the Curvatures.}
\label{subnuclear}

Our first task is to construct a set of curvature two-forms
representing the superalgebra $Osp(1|8)$. We begin with 
an explicit
$8\times 8$ matrix representation of the bosonic algebra
$Sp(8,{\bf R})$:
\be
\begin{array}{lll}
P_m=\divrt\g_m\otimes\s^+ &
M_{mn}=\half\g_{mn}\otimes \one &
K_m=-\divrt\g_m\otimes\s^- \\
E_{mn}=\divrt\g_{mn}\otimes\s^+ &
D=\one\otimes\s^3 &
F_{mn}=\divrt\g_{mn}\otimes\s^- \\
&A=-\g^5\otimes\s^3&\\
&V_m=-\g_m\otimes\one&\\
&Z_m=\g^5\g_m\otimes\s^3&
\end{array}
\label{rep}
\ee
The matrices $\s^\pm\equiv \s^1\pm i\s^2$ where $\s^1,\s^2$ and 
$\s^3$ are the usual Pauli matrices. Hence $\s^+$, $\s^-$ and
$\s^3$ are real and generate $SO(2,1)=Sl(2,{\bf R})$.
The matrices $\g_m$ are four real Dirac matrices satisfying
\be
\{\g_m,\g_n\}=2\neta_{mn}=2\diag(-1,1,1,1)_{mn},
\ee
and $\g^5\equiv\g^0\g^1\g^2\g^3$ 
is also real and satisfies $(\g^5)^2=-1$,
$\g^5=-(\g^5)^\top$.
Further,
$\g^{mn}\equiv\g^{[m}\g^{n]}=\frac{1}{2}(\g^m\g^n-\g^n\g^m)$.
It is easy to check that the matrices 
in\rf{rep}
close under commutation.
All matrices $M$ in\rf{rep} are real and since they satisfy 
$M^\top C+CM=0$ where $C=\g^0\otimes\s^1=-C^\top$ is the
direct product of the four and two-dimensional charge 
conjugation matrices $C_4=\g^0$ (where $C_4\g_m=-\g_m^\top C_4$) 
and $C_2=(-i\s^2)\s^3$ (where $C_2\s^\pm=(\s^\pm)^\top C_2$), 
they generate $Sp(8,{\bf R})$. 

The supersymmetric extension of the above matrix algebra 
is obtained
by introducing fermionic generators $Q_\a$ and $S_\a$ with
$\a=1,..,4$, both of which are Majorana spinors, possessing 
dilaton
weights $+1$ and $-1$, respectively 
(i.e., $[D,Q]=Q$, $[D,S]=-S$).
The commutation relations between fermionic and bosonic 
generators
(which may be obtained by requiring that 
$[\mbox{bose},$ $\{\mbox{fermi},\mbox{fermi}\}]$
and $[\mbox{fermi},\{\mbox{fermi},\mbox{fermi}\}]$ Jacobi
identities hold, or directly from the explicit
$SO(4,2)$--covariant
expressions for the $Osp(1|8)$ algebra given below)
imply that the explicit matrix representation\rf{decomp} 
acts on the
eight dimensional real $SO(2,4)$-spinor $Q_a = (S_\a, Q_\a)$: 
\be
\left[P_m,\left(\!\!
\begin{array}{c}S\\Q\end{array}\!\!\right) \right]
=-\sqrt{2}\left(\!\!
\begin{array}{c}\gamma_m Q\\ 0 \end{array}\!\! \right) .
\label{egg}\ee
To each of the generators
\be
T_A=\{P_m,E_{mn},Q,M_{mn},D,A,V_m,Z_m,S,K_m,F_{mn}\}
\ee
with (anti)commutation relations
\be
[T_A,T_B\}=f_{AB}{}^CT_C,  \phantom{]}
\ee 
we now associate a gauge field
one-form
\footnote{The gauge fields have mass dimensions
$[e_\m{}^m]=[E_\m{}^{mn}]=0$, $[\psi_\m]=1/2$, 
$[\o_\m{}^{mn}]=[b_\m]=[a_\m]=[v_\m{}^m]=[z_\m{}^m]=1$, 
$[\phi_\m]=3/2$, $[f_\m{}^m]=[F_\m{}^{mn}]=2$.}
\be
h^A=\{e^m,E^{mn},\psi,\o^{mn},b,a,v^m,z^m,\phi,f^m,F^{mn}\}
\ee
(where, to avoid overcounting,
for the generators $E_{mn}$, $M_{mn}$ and $F_{mn}$ the
indices take values $m<n$ only).
In particular we interpret the
$P_m$ gauge field
$e_\mu{}^m$ as the vierbein where 
$e_\mu{}^m e_{\nu m}=g_{\mu\nu}$.
The curvature two-form is then given by (for brevity we omit
wedge symbols)
\be
R=dh+hh=(dh^A-\half h^Ch^Bf_{BC}{}^A)T_A=R^AT_A\equiv\half
R_{\mu\nu}^AT_Adx^\mu dx^\nu,
\ee
and $h=h^A_\mu T_Adx^\mu$.

The fermionic terms in the bosonic curvatures may be obtained 
independently by a
method which we shall explain below. The results for the
curvatures are
\bea
\hspace{-.3cm}R(P)^m&=&
\begin{array}[t]{r}
de^m+\o^m{}_ne^n+2be^m-2E^m{}_nv^n-2\wt{E}^{m}{}_nz^n
         -\frac{1}{4\sqrt{2}}\psibar\g^m\psi
\end{array}
\label{curve}\\
\hspace{-.3cm}R(E)^{mn}&=&
\begin{array}[t]{r}
            dE^{mn}-2\o^{[m}{}_kE^{n]k}+2bE^{mn}                
            +2\wt{E}^{mn}a-4e^{[m}v^{n]}\\
            +2\e^{mnpq}e_pz_q
            +\frac{1}{4\sqrt{2}}\psibar\g^{mn}\psi
\end{array}\\
\hspace{-.3cm}R(Q)&=&
\begin{array}[t]{r}
d\psi+\left(-a\g^5+b+\frac{1}{4}\o^{mn}\g_{mn}-v^m\g_m
          +z^m\g^5\g_m\right)\psi\\
      +\left(\sqrt{2}e^m\g_m+\frac{1}{\sqrt{2}}E^{mn}\g_{mn}
      \right)\phi
\end{array}\\
\hspace{-.3cm}R(M)^{mn}&=&
\begin{array}[t]{r}
d\o^{mn}-\o^{[m}{}_k\o^{n]k}+4v^{[m}v^{n]}+4z^{[m}z^{n]}
             -8e^{[m}f^{n]}\\
          -8E^{[m}{}_kF^{n]k}+\frac{1}{2}\psibar\g^{mn}\phi
\end{array}\\
\hspace{-.3cm}R(D)&=&
\begin{array}[t]{r}
db-2e^mf_m-E^{mn}F_{mn}+\frac{1}{4}\psibar\phi
\end{array}\\
\hspace{-.3cm}R(A)&=&
\begin{array}[t]{r}
da-2v^mz_m-\wt{E}^{mn}F_{mn}
       +\frac{1}{4}\psibar \g^5\phi
\end{array}\\
\hspace{-.3cm}R(V)^m&=&
\begin{array}[t]{r}
dv^m+\o^m{}_nv^n+2z^ma+2E^m{}_nf^n-2F^m{}_ne^n
         +\frac{1}{4}\psibar\g^m\phi
\end{array}\\
\hspace{-.3cm}R(Z)^m&=&
\begin{array}[t]{r}
dz^m+\o^m{}_nz^n-2v^ma+2\wt{E}^{mn}f_n
         +2\wt{F}^{mn}e_n+\frac{1}{4}\psibar \g^5\g^m\phi
\end{array}\\ 
\hspace{-.3cm}R(S)&=&
\begin{array}[t]{r}
d\phi+\left(a\g^5-b+\frac{1}{4}\o^{mn}\g_{mn}-v^m\g_m
          -z^m\g^5\g_m\right)\phi\\
         +\left(-\sqrt{2}f^m\g_m+
\frac{1}{\sqrt{2}}F^{mn}\g_{mn}\right)\psi
\end{array}\\
\hspace{-.3cm}R(K)^m&=&
\begin{array}[t]{r}
df^m+\o^m{}_nf^n-2bf^m+2F^m{}_nv^n-2\wt{F}^{mn}z_n
         +\frac{1}{4\sqrt{2}}\phibar\g^m\phi
\end{array}\\            
\hspace{-.3cm}R(F)^{mn}&=&
\begin{array}[t]{r}
dF^{mn}-2\o^{[m}{}_kF^{n]k}-2bF^{mn}                
            -2\wt{F}^{mn}a+4f^{[m}v^{n]}\\
         +2\e^{mnpq}f_pz_q
            +\frac{1}{4\sqrt{2}}\phibar\g^{mn}\phi.
\end{array}
\label{curvF}
\eea
The duals\footnote{We remind the reader that in Minkowski space
$\wt{\wt{X}}^{{}_{{}_{{}_{{}_{{}_{\scriptstyle
mn}}}}}}=-X^{[mn]}$.} 
are defined by 
$\widetilde{X}^{mn}\equiv(1/2)\e^{mnpq}X_{pq}$, and the bars on
the Majorana fermions are defined by $\psibar=\psi^\top
C_4=\psi^\top\g^0$.
To fix the fermionic terms in the bosonic curvatures,
we used the Bianchi identities
\be
dR^A=-R^Ch^Bf_{BC}{}^A,\label{Bia}
\ee
Acting on any given curvature
with an exterior derivative $d$ yields terms of the 
form $dh^Ah^B$. 
Replacing $dh^A$ by $R^A$ minus its field--field terms, and
setting all curvatures $R^A=0$, all remaining
terms, which are cubic in fields,
must cancel amongst themselves as a 
consequence of the Jacobi identities. 
This fixes the coefficients of the fermionic terms in the
bosonic curvatures, for example the coefficient
of the $\psibar\g^m\psi$ term in
$R(P)^m$.
This cancellation of
combinations of terms cubic in fields will also be highly useful 
when we consider variations of actions.

It is interesting to recast the above results for the
$Osp(1|8)$ generators
and curvatures into an $SO(2,4)$--covariant form.
First, we define the supersymmetry generators
$ Q_{a} = a_{a} a$, 
with $[a_{a}, a_{b}] = - C_{ab}$ and $\{a,a\} = 1$ with indices
$a,b=1,...,8$.
We obtain
\bea
\hspace{-.3cm} \{ Q_{a} , Q_{b}\} &=&
\begin{array}[t]{r}
a_{(a} a_{b)} 
\end{array}
\eea
and use the Fierz identity
\bea
 \de^c_{(a}  \de^d_{b)} &=&
-\frac{1}{8} \Big\{  \G^7{}_{ab}\ \G^{7cd}
+ \frac{1}{2} \G^{MN}_{ab}\  \G_{MN}^{cd}  
+ \frac{1}{6} \G^{LMN}_{ab}\ \G_{LMN}{}^{cd} \Big\}
\eea
to rewrite this as
\bea
\hspace{-.3cm} \{ Q_a , Q_b \} &=&
\begin{array}[t]{r}
\frac{1}{8} \Big\{  \G^7_{ab}\ a^c \G^7_c{}^d a_d 
+ \frac{1}{2} \G^{MN}_{ab}\ a^c \G_{MN}{}_c{}^d a_d \\
+ \frac{1}{6} \G^{LMN}_{ab}\ a^c \G_{LMN}{}_c{}^d a_d \Big\}
\end{array}\\
 &\equiv&
\begin{array}[t]{r}
\frac{1}{4} \Big\{  \G^7_{ab}\ J_7  
+ \frac{1}{2} \G^{MN}_{ab}\ J_{MN}
+ \frac{1}{6} \G^{LMN}_{ab}\ J_{LMN} \Big\} \ .
\label{qq6}
\end{array}
\eea
The $SO(1,3)$-decomposition of the Gamma-matrices we use reads
\bea
\begin{array}{lll}
\G^m=-\g^m\otimes\s^3, &
\G^\oplus=\frac{1}{\sqrt{2}}\ \one \otimes\s^+, &
\G^\ominus=\frac{1}{\sqrt{2}}\ \one \otimes\s^-
\end{array} \\
\begin{array}{ll}
\{\Gamma^M,\Gamma^N\}=2\eta^{MN}={\rm diag}(--++++)^{MN}\ ,& 
\G^7 = -\g^5\otimes \s^3
\end{array}
\label{rep6}
\eea
where $\G^M = \G^M{}_a{}^b$ 
and we have chosen $\eta_{\oplus \ominus} = 1$.
We raise and lower indices as
follows: $a^a = C^{ab} a_b$, 
$\G^\ast{}^{ab} = \G^\ast{}^a{}_c C^{cb} 
= C^{ac} \G^\ast{}_c{}^b $, 
$\G^\ast{}_{ab} = \G^\ast{}_a{}^c C_{cb} = C_{ac} 
\G^\ast{}^c{}_b $,
$C^{ac} C_{cb} = \de^a_b$.  
$C^{ab} = ( \g^0 \otimes \sigma^1 )^{ab}$ is the $8\times8$
charge conjugation matrix introduced above. 
With these conventions, among the matrices 
$\G^\ast{}_{ab}$ we find
$\G^7,\G^{MN}$ and $\G^{MNP}$ symmetric
under interchange of $a$ and $b$, while
$C,\G^M,\G^{MNPQ}$ and $\G^{MNPQR}$ are antisymmetric.
Similarly, the real $4\times4$ matrices 
$\g^\ast{}_{\a\b}$ are split into
the symmetric $\g^m,\g^{mn}$ and the antisymmetric 
$C^4,\g^{mnp}$ and $\g^5$.
The remaining sectors of $Osp(1|8)$ now read: 
\bea
\hspace{-.3cm} [ J^\ast , Q_a ] &=&
\begin{array}[t]{r}
-\G^\ast{}_a{}^b Q_b 
\end{array}\label{quirk}\\
\hspace{-.3cm} [ J^7 , J^{MNP} ] &=&
\begin{array}[t]{r}
\frac{1}{3} \epsilon^{MNPRST} J_{RST}
\end{array}\\
\hspace{-.3cm} [ J^{MN} , J_{RS} ] &=&
\begin{array}[t]{r}
8\ \de^{[N}_{[R} J^{M]}{}_{S]}
\end{array}\\
\hspace{-.3cm} [ J^{MN} , J_{RST} ] &=&
\begin{array}[t]{r}
12\ \de^{[N}_{[R} J^{M]}{}_{ST]}
\end{array}\\
\hspace{-.3cm} [ J^{MNP} , J_{RST} ] &=&
\begin{array}[t]{r}
2\ \epsilon^{MNP}{}_{RST} J^7 - 36\ 
\de^{[M}_{[R}\de^{N}_{S} J^{P]}{}_{T]}
\end{array}\label{quark}
\eea
The bosonic generators decompose under $SO(1,3)$ as
\be
\begin{array}{lll}
P^m=\G^{\oplus m}&
M^{mn}=\frac{1}{2}\G^{mn} &
K^m=\G^{\ominus m}\\
E^{mn}=\G^{\oplus mn}&
D = \G^{\oplus\ominus} &
F^{mn}=\G^{\ominus mn} \\
&A = \G^7&\\
&V^m = \G^{\oplus \ominus m}&\\
&Z^m = -\frac{1}{3!} \e^{mnpq} \G_{npq}&
\end{array}
\label{decomp}
\ee
where the index structure of the matrices is $\G^\ast =  
\G^\ast{}_a{}^b$. 

In $SO(2,4)$-covariant language
the connection 1-forms are written as 
$h = h_7 J^7 + \half h_{MN} J^{MN}
+ \frac{1}{3!}  h_{MNP} J^{MNP} + \psi^a Q_a$, with 
$\psi^a = (\phi^\a , \psi^\a)$,
and the curvatures $R = dh + hh$
are given by
\bea 
\hspace{-.3cm} R &=&\hspace{-.9cm}
\begin{array}[t]{rl} 
&\hspace{-.3cm}
\begin{array}[t]{rl}
\Big\{ dh_7 + \frac{1}{36} \e^{MNPRST} h_{MNP}h_{RST} + 
\frac{1}{8} \psi^a \G_{ab} \psi^b & \!\!\!\!\!\Big\} J^7
\end{array}
\\ 
\hspace{.4cm}+ \ \ \half \!\!\!\!\!& 
\begin{array}[t]{rl}
\Big\{ dh_{MN}  + 2 h_{MK}h^K{}_{N} - h^{RS}{}_M h_{RSN}\\
+ \frac{1}{8} \psi^a \G_{MNab} \psi^b & \!\!\!\!\!\Big\} J^{MN}
\end{array}
\\ 
\hspace{.4cm}+ \ \ \frac{1}{6}\!\!\!\!\! &
\begin{array}[t]{rl}
\Big\{ dh_{MNP} + \frac{1}{3} \e_{MNPRST} h^{RST} h_7
+ 6 h_{M}{}^K h_{KNP}\\
+ \frac{1}{8} \psi^a \G_{MNPab} \psi^b & \!\!\!\!\!\Big\} J^{MNP}
\end{array}
\\ 
\hspace{.4cm}+ \!\!\!\!\!&
\begin{array}[t]{rl}
\Big\{ d\psi^a + h_7 \G^{7a}{}_b \psi^b + 
\frac{1}{2} h_{MN} \G^{MNa}{}_b \psi^b\\ +  
\frac{1}{6} h_{MNP} \G^{MNPa}{}_b \psi^b & \Big\} Q_a
\end{array}
\label{sixdecomp}
\end{array}
\eea
The gauge transformations $\de h = d\l + [h,\l]$
imply $\de R = [R,\l]$, i.e.
\bea
\hspace{-.3cm} \de R &=&\hspace{-.9cm}
\begin{array}[t]{rl}
&\hspace{-.3cm}
\begin{array}[t]{r}
\Big\{ \frac{1}{18} \e^{MNPRST} R_{MNP}\l_{RST} - 
\frac{1}{4} R^a \G^7{}_{ab} \l^b \Big\} J^7
\end{array}
\\ 
 \hspace{.4cm}+\ \ \half\!\!\!\!\!\! 
& \begin{array}[t]{r}
\Big\{ 4 R_{MK}\l^K{}_{N} - 2 R^{RS}{}_M \l_{RSN}
- \frac{1}{4} R^a \G_{MNab}\l^b \Big\} J^{MN} 
\end{array} 
\\ 
\hspace{.4cm}+ \ \ \frac{1}{6}\!\!\!\!\!\!
& 
\begin{array}[t]{rl}
\Big\{ - \frac{1}{3} \e_{MNPRST} R_7 \l^{RST} 
+ 6 \R_{M}{}^K \l_{KNP}& \\ 
+ \frac{1}{3} \e_{MNPRST} R^{RST} \l_7 
- 6 \R_{MN}{}^K \l_P{}^K & \\
- \frac{1}{4} R^a \G_{MNPab} \l^b & \Big\} J^{MNP} 
\end{array}
\\ 
\hspace{.4cm}+ \!\!\!\!\!\!&
\begin{array}[t]{rl}
\Big\{ R_7 \G^{7a}{}_b \l^b  
+ \frac{1}{2} R_{MN} \G^{MNa}{}_b \l^b   
+ \frac{1}{6} R_{MNP} \G^{MNPa}{}_b \l^b & \\ 
+ \l_7 \G^{7a}{}_b R^b + 
\frac{1}{2} \l_{MN} \G^{MNa}{}_b R^b +  
\frac{1}{6} \l_{MNP} \G^{MNPa}{}_b R^b & 
\!\!\!\!\!\Big\} Q_a
\end{array}
\label{sixdgaugetrans}
\end{array}
\eea
Using the explicit expressions for the 
$Osp(1|8)$-generators\rf{decomp}
it is straightforward to recover the corresponding 
$SO(1,3)$-decomposition
of the curvatures and gauge transformations. 

According to the algebraic program described in~\cite{PvN}
we split the symmetries generated by the $T_A$ into three
classes: ``sure'', ``automatic'' and ``unsure'' symmetries, 
which we now
explain. 
The action that we consider will be constructed
such that is it manifestly invariant under local Lorentz 
($M_{mn}$), dilation ($D$), axial ($A$) and 
general coordinate symmetries. These symmetries we call
``automatic'' symmetries. However, the usual gauge 
transformations 
\be
\delta h^A=d\e^A+\e^Ch^Bf_{BC}{}^A,\label{group}
\ee
for any gauge field $h^A$,
specialized to the case of $P^m$ gauge transformations 
(so that $\e^C$ is proportional to the parameter for
diffeomorphisms),
do not coincide with general coordinate transformations. 
Therefore, $P_m$ gauge transformations are no longer a symmetry
of the action, but are {\it replaced} by general coordinate
transformations. 
Hence, one no longer considers $P^m$ gauge transformations
on the left hand side of gauge commutators.
However, in order that the symmetry algebra
still closes, symmetries whose commutators {\it produce} 
$P_m$ gauge transformations must be modified. This is achieved
by imposing appropriate constraints on the curvatures. 
The symmetries $K_m$,
$F_{mn}$ and $S$ never produce a $P_m$ in commutators
because they have negative dilaton weights,
and are therefore unmodified. We call them ``sure''
symmetries since they must act on all fields according to the 
group law~(\ref{group}). 
(This is, of course, an assumption, to be justified by the
results.)
All constraints introduced into the
theory must be invariant under all symmetries, in particular
the sure and automatic
symmetries. The remaining gauge transformations for
the symmetries $V_m$, $Z_m$, $Q$ and
$E_{mn}$ will, in general,
be modified and are called ``unsure''
symmetries.

In more detail the way that constraints imply the modification
of the transformation rules of the unsure symmetries,
in contrast to the case of
$P_m$ gauge transformations which are simply 
exchanged for general
coordinate transformations, is as follows. The introduction of
constraints implies that by solving these constraints,
certain fields are expressed in terms of other fields; such
fields we call dependent fields.
In general, these dependent
fields no longer transform according to the 
group law~(\ref{group}), but rather, their transformation rules
are obtained by
varying their constituent independent fields according 
to~(\ref{group}) (the ``chain rule''). 
Of course, in the case of the sure and 
automatic symmetries which leave the constraints invariant,
the transformations of the dependent fields are left unmodified.

We shall determine these
constraints first dynamically and then kinematically: first we
shall require invariance of the action, and later we shall study
the implications for the relation between 
$P^m$ gauge transformations and general
coordinate transformations. As we shall show later, because
curvatures transform homogeneously under the group law, the
modifications in the transformation rules will contain terms
proportional to curvatures. However, further modifications,
beyond those implied by constraints, are,
in principle, not ruled out.

\section{The Set of Sure Constraints from Invariance of the
Action under Sure Symmetries.}\label{nuclear}

We now construct an affine action quadratic in
curvatures and invariant under
the sure symmetries $K_m$, $F_{mn}$ and $S$. 
By affine we mean that no vierbeins
are used to contract indices, but only constant Lorentz
tensors such as $\e^{\m\n\r\s}$, $\eta_{\m\n}$ and 
Dirac matrices. 
The most general
parity-even, Lorentz-invariant, 
dilaton-weight zero, mass dimension zero affine action
($S=\int_{M}\cal{L}$ for some four--manifold $M$) reads
\bea
-\cal{L}&=&\a_0\e_{mnpq}R(M)^{mn}R(M)^{pq}+\a_1R(A)R(D)+
           \a_2R(V)^mR(Z)_m\nonumber\\
       &&+\a_3\e_{mnpq}R(E)^{mn}R(F)^{pq}
          +\b\widebar{R(Q)}\g^5R(S).\label{action}
\eea
This action\footnote{The action is Hermitean and the curvatures are
real if one takes the reality condition for Majorana spinors
$\overline{\psi}=\psi^\top C_4=\psi^\dagger i\g^0$. We denote the
left hand side of the Minkowski action in\rf{action} by $-\calL$
to stress that we are using the metric $(-+++)$ rather than the
Euclidean notation of~\cite{Kaku1}. The sign $-\calL$ ensures
that the kinetic terms for the vierbein have the correct sign,
see, for example, reference~\cite{Fradkin}.}
is, of course, manifestly general coordinate
invariant since the 
integration measure $\e^{\m\n\r\s}$ is a
tensor density under general coordinate transformations.
The term $\a_2R(V)^mR(Z)_m$ is not chirally invariant
(i.e., w.r.t. the axial
gauge symmetry $A$), since $R(V)^m$ and $R(Z)^m$ undergo
an infinitesimal $SO(2)$ rotation, but all the other terms are
chirally invariant ($R(E)^{mn}$ and
$R(F)^{mn}$ transform into plus or minus their duals,
respectively). 
One could therefore consider setting 
the coefficient $\a_2=0$ at this point.
However, for now, we will leave $\a_2\neq0$, but requiring
invariance under sure $K^m$, $F^{mn}$ and $S$ symmetries
will in any case imply $\a_2=0$.
Since we are interested in a theory of gravity we
set $\a_0=1$. (In fact, no nontrivial solution exists for
$\a_0=0$). 

The requirement that the action in\rf{action} be invariant
under the sure symmetries yields
\be
\a_0=1,\a_1=-32,\a_2=0,\a_3=8,\b=-8,\label{coeffs}
\ee
as we now explain.
Under the sure $K^m$, $F^{mn}$ and $S$ symmetries, 
curvatures simply
rotate into curvatures according to the group law,
\be
\delta R^A=-R^C\e^Bf_{BC}{}^A.\label{hiatus}
\ee
One may therefore readily 
verify\footnote{To bring this result into the form quoted above
one needs
the Schouten identity 
\bea
\e_{mnpq}X_r\!\!&=&\!\!\e_{rnpq}X_m+\e_{mrpq}X_n+
\e_{mnrq}X_p+\e_{mnpr}X_q,
\nn\eea
from which the following most useful identity for any pair of
antisymmetric $4\times 4$ matrices $X$ and $Y$ may be derived
\bea
\wt{XY}^{mn}\equiv\frac{1}{2}\e^{mnpq}X_{pk}Y^k{}_q=
\frac{1}{2}[X\wt{Y}-\wt{Y}X]^{mn}\equiv
X_k{}^{[m}\wt{Y}^{n]k}.\nn
\eea}
that the variation of the 
action~(\ref{action}) under the sure symmetries $K_m$, $F_{mn}$
and $S_\a$ with parameters $\e^m$, $\e^{mn}$ and $\e^\a$,
respectively, is
\bea
\!\!\!\!\!\!\!\!-\delta_{K}{\cal L}\!\!\! &=&\hspace{-.4cm}
\begin{array}[t]{l}
-32R(P)_m\left[\widetilde{R}(M)^{mn}
  \!+\!\frac{\a_1}{16}R(A)\eta^{mn}\right]\e_n
  \!+\!\sqrt{2}\b\overline{R(Q)}\g^5\g^mR(Q)\e_m\\
+2\left[(\a_2+4\a_3)\widetilde{R}(E)^{mn}R(V)_m
            +(\a_2-4\a_3)R(E)^{mn}R(Z)_m\right]\e_n\\
  \end{array}\label{df}\\
\!\!\!\!\!\!\!\!-\delta_{F}{\cal L}\!\!\! &=&\!\hspace{-.3cm}
\begin{array}[t]{l}
        \widetilde{R}(\! E)_{mn}\! \left[
       (4\a_3\!-\!32)R(M)^m{}_k\e^{kn}\!\!
-\!(4\a_3\!+\!\a_1)(R(D)\e^{mn}\!\!-\! R(A)\widetilde{\e}^{mn})
                                                  \right]\\
-2\a_2R(P)_m\left[\e^{mn}R(Z)_n-\widetilde{\e}^{mn}R(V)_n
\right]
\end{array}\label{dF}\\
\!\!\!\!\!\!\!\!-\delta_{S}{\cal L}\!\! &=&\hspace{-.3cm}
\begin{array}[t]{l}\overline{R(Q)}\left[
              (\frac{\a_1}{4}-\b)(R(A)+\g^5R(D))
              +(\frac{\a_2}{4}+\b)\g^mR(Z)_m\right.\\
 \hspace{1.2cm}\left. \ \ 
+  (\frac{\a_2}{4}-\b)\g^5\g^mR(V)_m
         + (2+\frac{\b}{4})\widetilde{R}(M)^{mn}\g_{mn} 
                                       \right]\e\\
         +\overline{R(S)}\left[\sqrt{2}\b \g^5\g^mR(P)_m
              +(\frac{\b}{\sqrt{2}}+\frac{\a_3}{\sqrt{2}})
                    \g^{mn}\wt{R}(E)_{mn}\right]\e ,
\end{array}\label{dS}
\eea
We now must find constraints on the curvatures 
and values for the coefficients 
$\a_1$, $\a_2$, $\a_3$ and $\b$ such that the 
variations~(\ref{df})-(\ref{dS}) vanish. 

Consider first
$\delta_{K}{\cal L}$. The fermionic term must vanish
by itself and this is achieved via the constraint
\be
R(Q)=-\g^5{}_*R(Q)\label{const1},
\ee
where we denote the Hodge dual on curved indices
by a star ${}_*R_{\m\n}=(1/2)$ $e\e_{\m\n\r\s}R^{\r\s}$ and
$e=\det(e_\m{}^m)$. (Note that
$\overline{R(Q)}_{\m\n}\g^mR(Q)^{\m\n}$ vanishes for
Majorana spinors). In principle the sign in this constraint
is at this point arbitrary. However, we now impose the additional
requirement that all constraints should be solvable. By
solvable we mean that one can solve the constraint for some
field(s) algebraically by using the invertibility of the
vierbein. We shall
discuss this issue in detail later at the end of this
section, but at this point we observe the following:
In~\rf{const1} we have 12 constraints and in order that we can
solve for the 12 gamma-traceless components of
$\phi_\m-\frac{1}{4}\g_\m\g\cdot\phi$ instead of the
insufficient 4 gamma-trace components $\g\cdot\phi$, we need the
minus sign in\rf{const1}.
The definition\footnote{Recall also that in Minkowski space,
$\e_{\m\n\r\s}\e^{mnrs}=-e_\m{}^me_\n{}^ne_\r{}^re_\s{}^s+
\cdots$}
of $\e_{\m\n\r\s}$ which achieves this is such
that $\g_5\g_{\m\n}=\frac{1}{2} e \e_{\m\n\r\s}\g^{\r\s}$.

We turn now to the bosonic terms in
$\delta_{K}{\cal L}$. Solvable constraints contain a term  
linear in
curvatures and 
all terms in the constraint must have the same dilaton weight.
Hence the first and
last pair of terms in $\de_{K}\calL$ 
must cancel independently. Since
the constraints $R(M)^{mn}=0=R(A)$ are not solvable
(because $R(M)^{mn}$ has 36 components but one can at best
solve for the 16 components of $f_\m{}^m$ whilst $R(A)$
contains no terms with a vierbein), we are
forced to impose the constraint
\be
R(P)^m=0\label{const2}.
\ee
We note that both constraints~(\ref{const1}) and~(\ref{const2})
also appeared in the original conformal supergravity
case~\cite{Kaku,Kaku1}.
Let us now observe that the 
constraint~(\ref{const1}) should be invariant under the sure
symmetry $S$, while on the other hand we find that its variation
is given by
\be
\delta_S \left[R(Q)+\g^5{}_*R(Q)\right]
   =\frac{1}{\sqrt{2}}\left[R(E)^{mn}+{}_*\wt{R}(E)^{mn}\right]
     \g_{mn}\e.
\ee
We must therefore impose 
the additional (solvable) constraint
\be
R(E)^{mn}=-{}_*\widetilde{R}(E)^{mn}\label{const3}.
\ee
(This constraint is only solvable with the minus sign as given).
We then find that the remaining terms in $-\delta_{K}{\cal L}$
may be written as 
\be
2R(E)^{mn}[\a_2({}_*R(V)_m+R(Z)_m)+4\a_3({}_*R(V)_m-R(Z)_m)]\e_n
.
\ee
The curvature $R(E)^{mn}$ cannot be constrained to vanish as this
constraint would not be solvable. However the constraint 
\be
{}_*R(V)^m=R(Z)^m\label{const4}.
\ee
is solvable but is not solvable 
with the opposite sign.
Therefore we must take $\a_2=0$ (as predicted already by chiral
invariance) and impose the constraint\rf{const4} (we will find
later that $\a_3\neq0$).
One may verify that 
the set of constraints~(\ref{const1}), (\ref{const2}),
(\ref{const3}) and~(\ref{const4}) is invariant under
the sure and automatic symmetries.

Let us now consider $\delta_{F}\calL$. Again the first term
with $R(M)^{mn}$ should vanish by itself
because the traceless parts of $R(M)^{mn}$ and $R(E)^{mn}$
(``the Weyl parts'', see below) cannot be set to zero as these
constraints cannot be solved.
So we learn
that, as promised, 
$\a_3=8\neq0$. The constraint ${}_*R(A)=R(D)$ would kill
the terms with coefficients $4\a_3+\a_1$, but although solvable,
it produces the
wrong sign in~\rf{const1} when varied under the sure symmetry
$S$. (Further, under the sure symmetry $F_{mn}$, this constraint
${}_*R(A)=R(D)$ rotates into a double self-dual constraint for
$R(E)^{mn}$ instead of the required double anti-self-dual
constraint in~\rf{const3}. Under $K_m$, ${}_*R(A)=R(D)$ is
invariant.) Therefore
we take $\a_1=-4\a_3=-32$. 

Finally let us study $\delta_S \calL$. There is no solvable
constraint on $R(S)$ so that the last term must
be zero by itself.
This yields $\b=-\a_3=-8$. 
At this point the first two terms in~\rf{dS} cancel.
Also the term with $\wt{R}(M)^{mn}$ cancels since $\b=-8$
as in\rf{coeffs}.
We are therefore left only with
the third and fourth terms depending on $R(Z)^m$ and $R(V)^m$,
respectively, which may be rewritten using~(\ref{const1}) as
\be
\overline{R(Q)}\g^m
\left\{\frac{\a_2}{4}[R(Z)_m+{}_*R(V)_m]+
\b[R(Z)_m-{}_*R(V)_m]\right\}.
\ee
The term with $\b$ vanishes due to the constraint in~\rf{const4}.
Hence again we find $\a_2=0$. 
In summary, requiring invariance under the sure symmetries
$S$, $K_m$ and $F_{mn}$
has unambiguously led us to the following action and set of
solvable constraints
\bea
-\cal{L}&=&\e_{mnpq}R(M)^{mn}R(M)^{pq}-32R(A)R(D)\nonumber\\
       && +8\e_{mnpq}R(E)^{mn}R(F)^{pq}
          -8\widebar{R(Q)}\g^5R(S)
\label{S}
\eea
\bea
R(P)^m&=&0\label{c1}\\
R(E)^{mn}&=&-{}_*\widetilde{R}(E)^{mn}\label{c2}\\
R(Z)^m&=&{}_*R(V)^m\label{c3}\\
R(Q)&=&-\g^5{}_*R(Q).\label{c4}
\eea

The constraints on $R(P)^m$ and $R(Q)$ were found already in
conformal supergravity, and were not unexpected. The new
constraint in\rf{c2} is a direct consequence of the $R(Q)$ 
constraint in\rf{c4}. The constraint in\rf{c3} rotates into the
$R(Q)$ chiral self-dual constraint under $S$, hence it is
compatible with this constraint.
As the reader may verify from the above pages, not only are all
coefficients in the action fixed, but several reconfirmations of
our results were found in other sectors. The fact that one finds
over and over the same conditions on constraints
and parameters yields confidence in the results obtained so far.

We call the above set of constraints sure constraints since
they are necessary in order that the action is invariant under 
the sure symmetries. They are, however, not the maximal
set of solvable constraints that one could write down. We will
consider the maximal set of constraints in the next section,
but, as promised above, let us now discuss precisely what is
meant by the term ``solvable constraint''. 
A constraint is algebraically solvable only if it depends on the 
combination (vierbein)$\times$(field) 
and one can solve the constraint by expressing one or more
fields in terms of other fields. (The condition that the
constraint depends on the combination (vierbein)$\times$(field)
is of course only necessary
but not sufficient.)
For example, the constraint $R(P)^m=0$ is solved
for the spin connection $\o_\m{}^{mn}$ via
\bea
\o_{\m mn}&=&\frac{1}{2}(-\widehat{R}(P)_{mn\m}+
             \widehat{R}(P)_{\m mn}
           -\widehat{R}(P)_{\m nm});\nn \\ 
&& \widehat{R}(P)_{\m\n}{}^m
\equiv R(P;\o_\m{}^{mn}=0)_{\m\n}{}^m     
\label{omega}
\eea  
(To avoid confusion we alway write indices that
were originally curved to the left, for example we denote
$R(P)_{\m\n}{}^m e_a{}^\m e_b{}^\n \neta_{mc}$ by
$R(P)_{abc}$.)
In the case of conformal supergravity it was possible in this
way to find explicit expressions for dependent fields in terms
of the remaining independent fields only. However, notice now
that the above solution for the spin connection in~(\ref{omega}) 
is really only an expression for the spin connection in 
terms of other
dependent fields ($v^m$ and $z^m$).
In fact, in distinction
to the conformal supergravity case, 
our solutions to the constraints
only provide a set of coupled 
equations for the dependent fields.
In principle one could iterate this set of equations and it
is even possible that an iterated series solution could terminate
at some order. In any case, however, this is a new feature of
our extended conformal supergravity 
model which we shall discuss in
more depth below.

\section{The Maximal Set of Solvable Constraints.}
\label{Lorentz}

Let us now study the maximal set of constraints which are
solvable in the sense defined above. 
Clearly only curvatures depending explicitly on undifferentiated
vierbeins, namely the non-negative
dilaton weight curvatures $R(P)^m$, $R(E)^{mn}$, $R(Q)$,
$R(M)^{mn}$, $R(D)$, $R(V)^m$ and $R(Z)^m$, can be
solvably constrained, and conversely only the non-positive
weight gauge fields $f^m$, $F^{mn}$, $\phi$, $\o^{mn}$, $b$, 
$v^m$ and $z^m$ can appear in combination with a vierbein and
possibly be dependent.
To determine exactly which constraints are allowed we
must analyse which Lorentz irreducible pieces
of the curvatures can be constrained. 
The results are
summarized in figure~\ref{constraints} which we now explain in
more detail.

\setlength{\tabcolsep}{0.78\tabcolsep}
\begin{figure}[ht]
\begin{tabular}{|ccccccccccccccc|}
\hline
&&&&&&&&&&&&&&\\[-4mm]
$R(P)_{\m\n}{}^m $&=&\ul{24}&=&\ul{16}&+&$\wt{\ul{4}}$&+&\ul{4}&&&&&&\\
&&&&$\surd$&&$\surd$&&$\surd$&&&&&&\\[1mm]
\hline
&&&&&&&&&&&&&&\\[-4mm]
$R(E)_{\m\n}{}^{mn}$&=&
\ul{36}&=&$\wt{\ul{1}}$&+&\ul{10}
&+&$\wt{\ul{9}}$&+&\ul{9}&+&\ul{6}&+&\ul{1}\\
&&&&$\surd$&&$\times$&&$\surd$&&$\surd$&&$\surd$&&$\surd$\\[1mm]
\hline
&&&&&&&&&&&&&&\\[-4mm]
$R(Q)_{\m\n}$&=&
\ul{24}&=&\ul{8}&+&\ul{12}&+&\ul{4}&&&&&&\\
&&&&$\times$&&$\surd$&&$\surd$&&&&&&\\[1mm]
\hline
&&&&&&&&&&&&&&\\[-4mm]
$R(M)_{\m\n}{}^{mn}$&=&
\ul{36}&=&$\wt{\ul{1}}$&+&\ul{10}
&+&$\wt{\ul{9}}$&+&\ul{9}&+&\ul{6}&+&\ul{1}\\
&&&&$\times$&&$\times$&&$\times$&&$\surd$&&$\surd$&&$\surd$\\[1mm]
\hline
&&&&&&&&&&&&&&\\[-4mm]
$R(D)_{\m\n}$&=&\ul{6}&&&&&&&&&&&&\\
&&$\surd$&&&&&&&&&&&&\\[1mm]
\hline
&&&&&&&&&&&&&&\\[-4mm]
$R(V)_{\m\n}{}^m + {}_*R(Z)_{\m\n}{}^{m}$&=&
\ul{24}&=&\ul{16}&+&$\wt{\ul{4}}$&+&\ul{4}&&&&&&\\
&&&&$\surd$&&$\surd$&&$\surd$&&&&&&\\[1mm]
\hline
&&&&&&&&&&&&&&\\[-4mm]
$R(V)_{\m\n}{}^m - {}_*R(Z)_{\m\n}{}^{m}$&=&
\ul{24}&=&\ul{16}&+&$\wt{\ul{4}}$&+&\ul{4}&&&&&&\\
&&&&$\times$&&$\surd$&&$\surd$&&&&&&\\[1mm]
\hline
\end{tabular}
\caption{
Lorentz irreducible pieces of the ``solvable''
curvatures. The ticks ``$\surd$'' and crosses ``$\times$''
indicate those Lorentz irreducible pieces of
curvatures that may or may not,
respectively, be solvably constrained.\label{constraints}} 
\end{figure}

\hspace{-0.1cm}
The curvature $R(P)_{\m\n}{}^m$ has 24 independent components 
and can be decom\-posed into a trace
$R(P)_{\m\n}{}^\m=\underline{4}$, a totally antisymmetric part
$\e^{\m\n\r\s}R(P)_{\n\r\s}$ $=$ $\wt{\underline{4}}$ 
and the remaining
traceless piece ($\underline{16}$) all of which may
be constrained and solved for in terms of the 24 components of
$\o_{\m}{}^{mn}$. The constraint~(\ref{c1}) is clearly already
the maximal constraint possible for $R(P)_{\m\n}{}^m$.  
Note therefore,
in particular, that some combination 
of the Lorentz connection and the dilaton
connection is a dependent field, but whether the dilaton field or
the trace over the spin connection, or some combination of the
two, is the independent part is 
at this point not yet settled.
 
The curvatures $R(V)_{\m\n}{}^m$ and 
$R(Z)_{\m\n}{}^m$ have the same Lorentz decomposition 
as $R(P)_{\m\n}{}^m$.
The constraint $R(Z)^m={}_*R(V)^m$ 
in\rf{c3} yields 24 equations which can be solved 
by expressing the 24 components of $F_\m{}^{mn}$
in terms of other fields. Hence no further constraints involving
only $R(V)^m$ and $R(Z)^m$ are possible. 

The curvature $R(Q)_{\m\n}$ is a Majorana
spinor and hence has 24 real components which are decomposed into
a single gamma-trace $\g^\n$
$R(Q)_{\m\n}$ $-$ $\frac{1}{4}\g_\m
\g^{\a\b}$ $R(Q)_{\a\b}$ $=$ $\underline{12}$ (which vanishes
if traced a second time), 
a double gamma-trace
$\g^{\a\b}$ $R(Q)_{\a\b}$ $=$ $\underline{4}$ and a
gamma-traceless part (\underline{8}).
The constraint that the \underline{8} vanishes cannot be
solved
so it must be unconstrained. 
So far only the $\underline{12}$ is constrained by the chiral self-dual
constraint~(\ref{c4}). However the maximal constraint
\be
\g^\m R(Q)_{\m\n}=0\label{premc4}
\ee
was found to be necessary in the case of conformal supergravity
in order that the action be invariant under supersymmetry. 
It is solved for in
terms of all 16 components of $\phi_\m$ and is equivalent
to the sum of two constraints: the chiral self-dual constraint 
in~\rf{c4} and the double gamma-trace constraint
\be\g^{\m\n}R(Q)_{\m\n}=0.\label{Madonna}\ee
We shall later argue that also in our case 
the full constraint in\rf{premc4} must be imposed.

The curvature $R(E)_{\m\n}{}^{mn}$ has 36 components. 
Taking a single trace yields the 16 component Ricci
$R(E)_{\m\n}{}^{\n n}$ which may be further decomposed into
its antisymmetric (\underline{6}), trace (\underline{1}) and
symmetric traceless parts (\underline{9}) which may be solved
for in terms of the combinations $z_{[\m\n]}+{}_*v_{[\m\n]}$,
$v_\m{}^\m$ and
$v_{(\m\n)}-\frac{1}{4}g_{\m\n}v_\r{}^\r$,  
respectively. 
The remaining 20 trace free components 
decompose further into a piece antisymmetric in the interchange
of the first and last pair of indices ($\wt{\underline{9}}$),
a piece totally
antisymmetric in all four indices ($\wt{\underline{1}}$) and 
a \underline{10} which is the traceless piece 
symmetric in pairwise interchange
whose totally antisymmetric part ($\wt{\underline{1}}$) has
been subtracted out.
The $\underline{10}$ cannot be solved for but
the $\wt{\underline{9}}$ and $\wt{\underline{1}}$
may be solved for in terms of $z_{(\m\n)}-
\frac{1}{4}g_{\m\n}z_\r{}^\r$ and $z_\m{}^\m$ respectively.
If one now considers all combinations of $v_\m{}^m$ 
and $z_\m{}^m$ which can be solved for in all these constraints,
one sees that only the combination
$z_{[\m\n]}-{}_*v_{[\m\n]}$
does not appear in any constraint. There remains, of course, a
freedom to choose which combination of $z_{[mn]}$ and 
${}_*v_{[\m\n]}$ one takes to be dependent, but there will
always exist a second combination that remains independent. We
will investigate this freedom along with the freedom to
choose which combination of the dilaton field and trace
of the spin connection remains independent
in the following sections.

The double anti-self-dual constraint 
$R(E)^{mn}=-{}_*\wt{R}(E)^{mn}$
in~(\ref{c2})
contains 18 equations. There are two ways to obtain 18 by
combining the dimensions
\underline{1},
$\wt{\underline{1}}$,
\underline{6}, \underline{9}, $\wt{\underline{9}}$
and \underline{10}. 
One combination corresponds to \underline{1}, 
$\wt{\underline{1}}$, \underline{6} and \underline{10},
but since the constraint \underline{10} cannot
be solved it would be a disaster if\rf{c2} would correspond
to this combination of constraints. Fortunately,
the sign in\rf{c2} is precisely such that it corresponds to the
solvable combination of constraints with dimensions
\underline{9} and $\wt{\underline{9}}$.
Our discussion implies that further solvable constraints
on $R(E)^{mn}$ are possible, namely those with dimensions
\underline{1}, $\wt{\underline{1}}$ and \underline{6},
which can be written explicitly as follows
\bea
R(E)_{\n[\m m]}{}^\n&=&0\label{shalom}\\
R(E)_{\m\n}{}^{\n\m}&=&0\label{mimi}\\
\e^{\m\n\r\s}R(E)_{\m\n\r\s}&=&0.\label{mia}
\eea
Note that the constraints\rf{shalom}-(\ref{mia}) also follow
as a consequnce of the constraint\rf{Madonna}
by requiring invariance under the sure symmetry $S$.
(The variation of $\g^{\m\n}R(Q)_{\m\n}=0$ leads to 16
conditions, 8 of which are represented
by\rf{shalom}-(\ref{mia})).

It is easy to check that the complete set of solvable
constraints for $R(E)^{mn}$,
$R(Q)$, $R(V)^m$ and $R(Z)^m$
are invariant under the sure and
automatic symmetries.
We shall later see that the additional constraints\rf{Madonna},
(\ref{shalom}), (\ref{mimi}) and\rf{mia} (i.e., those
constraints which were not required for the invariance of the
action under the sure symmetries) must be imposed in order to
obtain invariance under $V_m$ and $Z_m$ symmetries. 

The Riemann tensor $R(M)_{\m\n}{}^{mn}$ has the same
decomposition as $R(E)_{\m\n}{}^{mn}$, however, only the 16
components of the Ricci tensor $R(M)_{\m\n}{}^{\n n}$ 
can be constrained and solved for in terms of the field
$f_\m{}^m$. 
Therefore, a constraint on the Ricci tensor
along with all constraints given above, represents the
{\it maximal}
set of solvable constraints.
The construction of such a Ricci\footnote{To avoid confusion,
note that, in contrast to torsionless Riemannian 
general relativity, the Ricci tensor
here has also an antisymmetric part (a \underline{6}) and is,
therefore, a completely general two index tensor.}
constraint is the subject of the next section. 

\section{The Complete Ricci Constraint}
\label{Myrtle}

The construction of a constraint on the Ricci curvature,
which is solvable in terms of
$f_\m{}^m$ and is invariant under the 
sure symmetries, is somewhat subtle. 
Moreover we observe that the 
antisymmetric part of $f_{\m\n}$ (a \underline{6})
occurs not only in the \underline{6}
of $R(M)_{\m\n}{}^{mn}$ but also in the \underline{6}
of $R(D)_{\m\n}$. Hence, it is not clear at this point 
whether one should constrain the \underline{6} of
$R(M)_{\m\n}{}^{mn}$, or $R(D)_{\m\n}$, or perhaps a linear
combination. In
fact, as we shall argue, both should be constrained. Since 
the same situation occurred in conformal supergravity, we
temporarily digress to the latter model.
In that model we shall obtain a useful
new interpretation of the maximal solvable 
$R(M)_{\m\n}{}^{mn}$
constraint which we will generalize to our model of
extended\footnote{The terminology ``extended'' as used
here, should, of
course, not be confused with the more common usage referring to 
$N\geq1$ supersymmetries.} 
conformal supergravity.

\subsection*{Intermezzo: Conformal Supergravity.}

Conformal supergravity is the gauge theory of the 
superconformal algebra $SU(2,2|1)$. An explicit
(reducible) representation of the bosonic conformal
algebra $SU(2,2)\subset Sp(8)$ is obtained from
the bosonic representation in~(\ref{rep}) by dropping the
extensions $E_{mn}$, $V_m$, $Z_m$ and $F_{mn}$. 
Its supersymmetric extension $SU(2,2|1)$ is summarized 
by the following representation in terms of curvature two-forms
\bea
\hspace{-.8cm}R(P)^m&=&
de^m+\o^m{}_ne^n+2be^m-\frac{1}{4\sqrt{2}}\psibar\g^m\psi\\
\hspace{-.8cm}R(Q)&=&
d\psi+(3a\g^5+b+\frac{1}{4}\o^{mn}\g_{mn})\psi
          +\sqrt{2}e^m\g_m\phi\\
\hspace{-.8cm}R(M)^{mn}&=&
d\o^{mn}-\o^{[m}{}_k\o^{n]k}-8e^{[m}f^{n]}\nonumber
         +\frac{1}{2}\psibar\g^{mn}\phi\\
\hspace{-.8cm}R(D)&=&
db-2e^mf_m+\frac{1}{4}\psibar\phi\\
\hspace{-.8cm}R(A)&=&
da+\frac{1}{4}\psibar \g^5\phi\\
\hspace{-.8cm}R(S)&=&
d\phi+(-3a\g^5-b+\frac{1}{4}\o^{mn}\g_{mn})\phi
-\sqrt{2}f^m\g_m\psi\\
\hspace{-.8cm}R(K)^m&=&
df^m+\o^m{}_nf^n-2bf^m+\frac{1}{4\sqrt{2}}\phibar\g^m\phi.    
\eea
Note that the superalgebra $SU(2,2|1)$ is not
a subalgebra of $Osp(1|8)$. This is most clearly seen
by analyzing their embedding in $Osp(2|8)$~\cite{Fradlin}: 
let the oscillators
$a_{A} = (a^{K}, \ol a_{K}, a, \ol a)$ have the (anti)commutation
relations $[a^{K} ,\ol a_{L} ] = \de^{K}_{L}$, $\{a,\ol a\}= 1$.
Here $\ol a_{K} = \eta_{K \dot L} a^{\dot L}$ is up to the
$SU(2,2)$-metric $\eta_{K \dot L}$ the complex conjugate
of $a_{K}$. A real $Sp(8)$-spinor is represented by the complex 
pair $(a^{K}, \ol a_{K}) = a_a$.
$Osp(2|8)$ has a total of 16 real supersymmetry charges, namely
the two $Sp(8)$-multiplets $Q^{(+)}_a = a_a a = ( a^{K} a , \ol 
a_{K} a)$
and  $Q^{(-)}_a = a_a \ol a = ( a^{K}\ol a , \ol a_{K}\ol a)$.
The subalgebra $SU(2,2|1)$ is obtained by selecting 
the supercharges $Q^{K} =  a^{K} \ol a$ and  $\ol Q_{K} =  \ol 
a_{K} a$,
while $Osp(1|8)$ contains eight different supercharges, namely
$Q_a = a_a (a+\ol a)/\sqrt{2}$.
As a consequence, the $Q_a$ $Q_b$ anticommutator produces,
among other things, an
$E_{mn}$ generator. The $SU(2,2|1)$-curvatures thus are not 
obtained by
setting the new fields $E^{mn}$, $v^m$, $z^m$ and $F^{mn}$
to zero. 
Note, however, that one only needs to give the the  
$a \g^5\psi$ and $a \g^5\phi$ terms in
the $R(Q)$ and $R(S)$ curvatures, respectively, 
an additional 
factor $-3$. One may check that the Fierz identities
required to obtain the Bianchi identity for the curvatures
hold only with the correct factor $-3$ given above.
In the oscillator representation one readily sees this factor:
the bosonic generators of $SU(2,2|1)$ are given by 
$J^K{}_L = \half \{ a^K , \ol a_L \} - \frac{1}{8} \de^K_L 
\{a^N , \ol a_N\}$
and $J = \half \{ a^K ,\ol a_K \} = \half [ a, \ol a]$ (which 
implies a
nontrivial trace condition on the total Hilbert space) and
hence
\be
\{Q^K , \ol Q_L\} 
\ = \  \half \{ a^K ,\ol a_L \}  -  \half \de^K_L [ a, \ol a]
\ = \  J^K{}_L - \frac{3}{4}\de^K_L J \ ,
\ee
while for $Osp(1|8)$ we obtain
\be
\{Q^K, \ol Q_L\} \ = \  \half \{ a^K , \ol a_L \} 
\ = \  J^K{}_L + \frac{1}{4} \de^K_L J \ ,
\ee
where we have defined $J = \half \{ a^K ,\ol a_K \}$ in the 
same fashion.  
Apart from this factor, and of course the generators 
$J^{KL}= a^{(K} a^{L)}$ and $\ol J_{KL} = \ol a_{(K} 
\ol a_{L)}$, 
the two algebras are identical.

One can give an explicit 5 $\times$ 5 matrix
representation~\cite{Kaku} of
$SU(2,2|1)$, with the axial generator $A$ represented by a
supertraceless diagonal matrix with entries proportional to
$(1,1,1,1,-4)$.

Again one may write down the most general parity even,
dilaton weight zero, affine action
and fix the coefficients and sure constraints by
requiring invariance with respect to the sure symmetries 
$K_m$ and $S$. The results 
are\footnote{Observe that the term $32R(A)R(D)$ appears with an
opposite sign from the $Osp(1|8)$ case. To study the sign of the
kinetic terms of the axial $a_\m$ and vierbein $e_\m{}^m$
fields, one must substitute the leading terms of the solution
of the constraint\rf{sfeqn} below
for the conformal boost gauge field $f_\m{}^m$ into\rf{sS}.
The result is
\bea
-\calL_{\rm
Kin}&=&-2[R(\o)_{\m\n}R(\o)^{\n\m}-\frac{1}{3}R(\o)^2]
+24R(A)_{\m\n}
R(A)^{\m\n},\nn
\eea
where all terms have the 
required sign and
$R(\o)_{\m\n}\equiv R(\o)_{\a\m\n}{}^\a$ and 
$R(\o)\equiv R(\o)_\m{}^\m$
are the Ricci and scalar curvatures, respectively,
of the usual Riemann curvature
$R(\o)_{\m\n}{}^{mn}=2\d_{[\m}\o_{\n]}{}^{mn}
+2\o_{[\m}{}^{k[m}\o_{\n]}{}^{n]}{}_k$.
Note, however, that the analagous calculation
(using now the constraint\rf{feqn} derived below) 
in the same sector of the
$Osp(1|8)$ model yields
\bea
-\calL_{\rm
Kin}&=&-2[R(\o)_{\m\n}R(\o)^{\n\m}-\frac{1}{3}R(\o)^2]
-8R(A)_{\m\n}
R(A)^{\m\n},\nn
\eea
so that the kinetic terms of the axial gauge field now
appear with the opposite (unphysical) sign.
}
\be
-{\cal L}=\e_{mnpq}R(M)^{mn}R(M)^{pq}+32 R(A)R(D)
         -8\overline{R(Q)}\g^5R(S),
\label{sS}
\ee
\bea
R(P)^m&=&0\label{sc0}\\
R(Q)&=&-\g^5{}_*R(Q),\label{sc1}\\
R(A)&=&{}_*R(D).\label{sc2}
\eea 
These constraints are themselves invariant under the sure
symmetries but do not represent a maximal set of constraints.
In the early work of Kaku et. al.~\cite{Kaku,Kaku1}, the
constraint~(\ref{sc2}) was not yet known (but rather was
discovered later in the algebraic approach in~\cite{PvN}).
Furthermore, instead of the affine action in\rf{sS} a 
non-affine action\footnote{Note that $R(A)={}_*R(D)$ implies
$R(D)=-{}_*R(A)$.}
\be
\calL_{\rm non-affine}=\calL +64R(A)(R(D)+{}_*R(A)),
\label{snongeom}
\ee
was employed which is clearly equivalent to~(\ref{sS})
since it differs only by the constraint~(\ref{sc2}). 
Consequently, also\rf{snongeom} is invariant under $K$ and $S$
symmetry.
The
non-geometric action is interesting for two reasons:
firstly, it is invariant under the sure symmetries
without having to use~(\ref{sc2}), and that was how it was
found in~\cite{Kaku1}. Second, if one considers
the algebraic field equation for the $f^m$ field in the
non-affine action
\be
0=\wt{R}(M)^{mn}e_n+2R(A)e^m
-\frac{1}{2\sqrt{2}}
\psibar \g^5\g^mR(Q),
\label{sfeqn}
\ee
it is easy to check that it is invariant under the
sure symmetries. Therefore, (\ref{sfeqn}) also represents 
a possible constraint on $R(M)^{mn}$ which may be solved
in terms of $f^m$. Imposing~(\ref{sfeqn}) as a constraint
implies, of course, that the non-geometric action is invariant 
under arbitrary variations of the field $f^m$
(the so-called 1.5 order formalism in supergravity~\cite{PR}). 
Rather than imposing\rf{sfeqn} as a field equation, it would be
preferable to consider it as a constraint, on a par 
with\rf{sc0}-(\ref{sc2}). Therefore, in~\cite{PvN} the
action\rf{sS} was chosen and\rf{sfeqn} imposed. 
Of course, 
in the affine action\rf{sS}
the field equation for $f^m$ no longer coincides with the
constraint~(\ref{sfeqn}).

The action~(\ref{sS}) (and clearly also~(\ref{snongeom}))
is obviously Lorentz ($M_{mn}$), dilation ($D$), axial ($A$)
and general coordinate invariant so one only needs to check
invariance under the unsure symmetry $Q$. The fields
$e^m$, $\psi$, $b$ and $a$ transform under $Q$ according to the 
group law (i.e. as ordinary gauge fields), but the dependent
fields $\o^{mn}$, $\phi$ and $f^m$ get extra transformations 
in order that the constraints remain valid. It is
therefore convenient to work with the non-affine action
(which is, at this point, completely equivalent to the affine
action because the constraint $R(A)={}_*R(D)$ holds) 
since one therefore need not calculate the extra transformations
of the field $f^m$. In~\cite{Kaku1} it is shown that
the theory is supersymmetry ($Q$) invariant if one extends the
constraint~(\ref{sc1}) to read
\be
\g^\m R(Q)_{\m\n}=0.\label{sc4}
\ee
The constraints~(\ref{sc0}), (\ref{sfeqn})
and~(\ref{sc4}) represent the maximal set of solvable
constraints in this model.

Finally, let us complete this section by showing that 
the the constraint~(\ref{sc2}) follows as a consequence of
the constraints~(\ref{sfeqn}) and~(\ref{sc4}).  
The Bianchi identity for the constraint~(\ref{sc0})
written in the form~(\ref{Bia}) yields a new constraint
\be
0=dR(P)^m=R(M)^{mn}e_n+2R(D)e^m+\frac{1}{2\sqrt{2}}\psibar
\g^mR(Q).\label{sbianchi}
\ee
Although the form of the two constraints~(\ref{sfeqn}) 
and~(\ref{sbianchi}) appears similar, in fact they are 
inequivalent,
since the former is a constraint on the Ricci ($\underline{1}$, 
$\underline{6}$ and $\underline{9}$) of $R(M)^{mn}$
whereas the Bianchi identity~(\ref{sbianchi}) constrains 
the $\wt{\underline{1}}$, $\underline{6}$ and 
$\wt{\underline{9}}$. There is no
contradiction with the claim above that only the Ricci 
part of $R(M)_{\m\n}{}^{mn}$ could be solvably constrained since
the Bianchi identity is an {\it identity} so 
that~(\ref{sbianchi}) holds simply because the 
dependent field $\o_\m{}^{mn}$ is the solution to $R(P)^m=0$.
The two constraints overlap on the $\underline{6}$ of 
$R(M)_{\m\n}{}^{mn}$ and using the constraint $\g^\m
R(Q)_{\m\n}=0$ in~\rf{sc4} they may be rewritten as the
following constraints on the \underline{6}
\bea
0&=&R(M)_{\r[\m\n]}{}^\r+2_*R(A)_{\m\n}+
\frac{1}{4\sqrt 2}\overline{R(Q)}_{\m\n}\g\cdot\psi
\hspace{.3cm}\{\mbox{$f^m$ equation}\}\\
0&=&R(M)_{\r[\m\n]}{}^\r-2R(D)_{\m\n}
+\frac{1}{4\sqrt{2}}\overline{R(Q)}_{\m\n}\g\cdot\psi
\hspace{.3cm}\{\mbox{Bianchi}\}.
\eea
The difference is clearly given by~(\ref{sc2}).

The reader may feel that the way we have arrived at the 
constraint in~\rf{sfeqn} was rather indirect (a field equation
in one model was turned into a constraint of another model),
but a completely kinematical approach yields all these
constraints directly. Requiring compatibility
between $P_m$ gauge transformations and general coordinate
transformations as discussed in section~\ref{subnuclear}
leads directly to the constraints\rf{sc0}, (\ref{sc1})
and\rf{sfeqn} from which\rf{sc1} follows
in the manner indicated~\cite{PvN}. 

\subsection*{Back to $Osp(1|8)$.}

Motivated by the above discussion, we consider a one
parameter ($\a$) family\footnote{Also in the conformal
supergravity case, without altering any of the conclusions
outlined in the above intermezzo, one can consider a one
parameter family of additional non-affine terms in the action 
that vanish modulo the constraint
$R(D)=-{}*R(A)$, namely $0=64(1-2\a)R(D)R(A)
64\a R(D){}_*R(D)
64(1-\a)R(A){}_*R(A)$, although in~\cite{Kaku1} only the case
with no $R(D){}_*R(D)$ term ($\a=0$)  was considered.} of 
non-affine
actions differing from~(\ref{S}) only by terms that vanish due to
the constraint $R(V)^m=-{}_*R(Z)^m$
\be
{\cal L}_{\rm non-affine}={\cal L}
\begin{array}[t]{l}
         +32(1-2\a)R(V)^mR(Z)_m
         +32\a R(V)^m{}_*R(V)_m\\
         +32(1-\a)R(Z)^m{}_*R(Z)_m.
     \end{array}\label{Ss}
\ee
One may consider more general non-affine actions equivalent
to the action~(\ref{S}) but the requirement that the 
$f^m$ field equation be invariant under the sure symmetries
leads one to~(\ref{Ss}).
We propose the $f^m$ field equation of these non-affine
actions as the \underline{16} 
constraint for $R(M)^{mn}$ which reads
\bea
0&=&\wt{R}(M)^{mn}e_n+2R(A)e^m
-\frac{1}{2\sqrt{2}}
\psibar \g^5\g^mR(Q)
-2\wt{R}(E)^{mn}v_n\nonumber\\&&-2\wt{E}^{mn}R(V)_n
+2R(E)^{mn}z_n+2E^{mn}R(Z)_n,\label{feqn}
\eea
where {\it after} varying the non-affine action in~\rf{sS} under
$f^m\rightarrow f^m+\de f^m$, we imposed the constraint
$R(V)^m=-{}*R(Z)^m$.
The new constraint in~\rf{feqn} 
is invariant under the sure and
automatic symmetries and is 
to be compared with the following Bianchi identity
\bea
0=dR(P)^m&=&R(M)^{mn}e_n+2R(D)e^m+\frac{1}{2\sqrt{2}}
\psibar\g^mR(Q)-2R(E)^{mn}v_n\nonumber\\&&
+2E^{mn}R(V)_n
-2\wt{R}(E)^{mn}z_n+2\wt{E}^{mn}R(Z)_n.\label{Bianid}
\eea
The constraint~(\ref{feqn}) may also be found by writing down
the most general three form constraint 
$0=\wt{R}(M)^{mn}e_n+\ldots$ with fixed parity and dilaton
weight two, 
and fixing the coefficients by requiring invariance under the
sure symmetries. We note also that, just like the Bianchi
identity, to achieve invariance of the
constraint~(\ref{feqn}) under sure symmetries, 
one needs only use that $R(P)^m=0$.

Again, the constraints~(\ref{feqn}) and~(\ref{Bianid})
overlap on the \underline{6} of $R(M)_{\m\n}{}^{mn}$
so let us compute the difference of these two constraints
which will, as a consequence, lead to another useful constraint.

The \underline{6} of~(\ref{feqn}) and~(\ref{Bianid})
is given, respectively, by
\bea
0&=&{}_*\wt{R}(M)_{\r[\m\n]\r^\prime}g^{\r\r^\prime}
-2{}_*R(A)_{\m\n}+
\frac{1}{4\sqrt{2}}\psibar\cdot\g R(Q)_{\m\n}\nn\\
\!&&\hspace{-.2cm}-R(E)_{\m\n\r\s}v^{\r\s}
+2\wt{E}^{\r\s}{}_{[\m}R(Z)_{\n]\r\s}
-{}_*R(E)_{\m\n\r\s}z^{\r\s}
+2E^{\r\s}{}_{[\m}R(V)_{\n]\r\s}\nn\\ \label{das}\\
0&=&{}_*R(M)_{\r[\m\n]}{}^\r-
2{}_*R(D)_{\m\n}-\frac{1}{4\sqrt{2}}
\psibar\cdot\g {}_*R(Q)_{\m\n}\nn\\
\!&&\hspace{-.3cm}+{}_*R(E)_{\m\n\r\s}v^{\r\s}
-2E^{\r\s}{}_{[\m}R(Z)_{\n]\r\s}
-R(E)_{\m\n\r\s}z^{\r\s}
+2\wt{E}^{\r\s}{}_{[\m}R(V)_{\n]\r\s}.\nn \\
\label{it}
\eea
To obtain\rf{das} and\rf{it} from\rf{feqn} and\rf{Bianid},
respectively, along with the sure constraints\rf{c2}, (\ref{c3})
and\rf{c4}, 
we used the following cyclicity relations:
\be
R(E)_{[\m\n\r]\s}=0\label{Ecyc}
\ee 
which constrains the $\wt{\underline{1}}$, $\wt{\underline{9}}$
and $\underline{6}$ of $R(E)_{\m\n}{}^{mn}$ to zero as
in\rf{mia},
half of\rf{c2} and\rf{shalom}, respectively,
\be
{}_*R(E)_{[\m\n\r]\s}=0\label{Ecyc1}
\ee 
which constrains the $\underline{1}$, $\underline{9}$
and $\underline{6}$ of $R(E)_{\m\n}{}^{mn}$ to zero as
in\rf{mimi},
the other half of\rf{c2} and again\rf{shalom}, respectively,
and 
\be
\g_{[\m} R(Q)_{\r\s]}=0\label{cycle2}
\ee
which follows from the constraint~(\ref{premc4}).
Dualizing\rf{das} in the indices $\m$ and $\n$ and then
adding this result to\rf{it} yields (dividing by an
overall factor $2$)
\bea
0&=&R(A)_{\m\n}-{}_*R(D)_{\m\n}
+\wt{E}^{\r\s}{}_{[\m}R(V)_{\n]\r\s}
-E^{\r\s}{}_{[\m}R(Z)_{\n]\r\s}
\nn\\&&\hspace{-.4cm}+\frac{1}{2}e\e_{\m\n\a\b}
\left[E^{\r\s\a}R(V)^{\b}{}_{\r\s}
+\wt{E}^{\r\s\a}R(Z)^{\b}{}_{\r\s}\right].\label{six}
\eea
To see that the higher order terms in\rf{six} 
do not cancel one can use $R(V)^m=-{}_*R(Z)^m$ to write the 
result in terms of $R(Z)^m$ only.

{\it In summary}, we have found the following maximal set of
solvable constraints
\bea
0&=&R(P)_{\m\n}{}^m\label{mc1}\\
0&=&R(E)_{\r[\m \n]}{}^\r\label{mc2}\\
0&=&R(E)_{\m\n}{}^{\n\m}\\
0&=&\e^{\m\n\r\s}R(E)_{\m\n\r\s}\\
0&=&R(E)_{\m\n}{}^{mn}+{}_*\wt{R}(E)_{\m\n}{}^{mn}\label{mc25}\\
0&=&R(Z)_{\m\n}{}^m-{}_*R(V)_{\m\n}{}^m\label{mc3}\\
0&=&\g^\m R(Q)_{\m\n}\label{mc4}
\eea
\vspace{-.6cm}
\bea
0\!\!&=&\!\!
R(M)_{\r\m\n}{}^\r-\frac{1}{2}g_{\m\n}R(M)_{\r\s}{}^{\s\r}
+2{}_*R(A)_{\m\n}+\frac{1}{2\sqrt{2}}\overline{R(Q)}_{\r\n}
\g_\m\psi^\r\nn\\\!\!\!&&
\!\!-2{}_*R(E)_{\r\n\s\m}z^{\r\s}
-2R(E)_{\r\n\s\m}v^{\r\s}+2R(V)_{\r\n\s}E^{\r\s}{}_\m
+2R(Z)_{\r\n\s}\wt{E}^{\r\s}{}_\m.\nn\\&&\label{mc5}
\eea
All further constraints must follow from this set, either
algebraically or, for example, by Bianchi identities.

\section{Invariance of the Action under $V_m$ and $Z_m$
Symmetries at the Linearized Level.}
\label{Linear}

At this point we have found the maximal set of
solvable constraints and an action, both of which are invariant
under sure and automatic symmetries. In principle,
therefore, we should now simply vary the action with respect to
the remaining unsure symmetries taking into account that 
the constraints endow the dependent fields with extra
transformations over and above the usual group law
gauge transformations. However, 
unlike the conformal supergravity case, 
where one had explicit solutions for the dependent fields in
terms of independent fields so that the above program could be
(and was~\cite{Kaku1}) directly carried out, we must now grapple
with the fact that we only have a coupled set of equations
for the dependent fields whose iterative solution, in general,
is an infinite series in independent fields.

Of course one may argue that to investigate the invariance of
the action under the remaining symmetries one needs only the 
extra transformations of dependent fields rather than 
explicit solutions for the fields themselves. However again 
the same problem arises, namely that the constraints provide only
a coupled set of equations for the extra transformations of 
dependent fields. In order to calculate further, 
we make a consistent expansion in the number of fields
and study the model in the lowest order in this expansion.

Consider, for example, the constraint $R(P)^m=0$ and some
unsure symmetry which we denote by $\de$. On
independent fields $\de$ acts simply as a gauge transformation

\be
\de h^A_{\rm Indept.}=d \e^A+\e^C h^Bf_{BC}{}^A
=\de_{\rm Group}h^A_{\rm Indept.}. 
\ee
However acting on dependent fields we have
\be
\de h^A_{\rm Dept.}=d
\e^A+\e^C h^Bf_{BC}{}^A+\what{\de}h^A_{\rm Dept.}
\equiv \de_{\rm Group}h^A_{\rm Dept.}+
\what{\de}h^A_{\rm Dept.}, 
\ee
where the extra transformations 
$\what{\de}h^A_{\rm Dept.}$ are determined by
requiring that the constraints are invariant under the unsure
symmetry $\de$, for example
\bea
0=\de R(P)^m&=&\de_{\rm Group}R(P)^m+\what{\de}R(P)^m\nonumber\\
&=&\de_{\rm Group}R(P)^m+\dehat\o^{mn}e_n
-2E^{mn}\dehat v_n-2\wt{E}^{mn}\dehat z_n.\label{solve}
\eea
Note that in\rf{solve}, the extra transformations of 
three\footnote{For simplicity we have given\rf{solve} for the
case in which the dilation gauge field $b$ remains
independent.} dependent fields appear, as opposed to conformal
supergravity where only $\dehat \o^{mn}$ was present.
One may write down similar expressions for all other
constraints which {\it in principle} 
uniquely determine all extra transformations of the 
dependent fields. 

In practice, to write down a solution for the 
extra transformations of the dependent fields, we solve the set
of coupled equations for the extra transformations given by
requiring invariance of the constraints iteratively.
Namely, we make an expansion order by order in the number of
independent fields, where one counts the vierbein as a
Kronecker delta (i.e. field number zero).
In this section, we show that the action is invariant under
$V_m$ and $Z_m$ symmetries at the leading order of this
expansion.
 
For example, in~(\ref{solve}) one first ignores the terms 
$-2E^{mn}\dehat v_n-2\wt{E}^{mn}\dehat z_n$ since they are 
next to leading order and solves for the lowest order
contribution to $\dehat \o^{mn}$ which is given by
\be
\dehat\o_{\m mn}=\frac{1}{2}(-\de_{\rm Group} R(P)_{mn\m}+
             \de_{\rm Group} R(P)_{\m mn}
           -\de_{\rm Group}
 R(P)_{\m nm})+\cdots.
\ee
One can proceed similarly for all other constraints and then
insert the leading order results in the next order contributions
thereby generating expressions of the form
\be
\dehat h^A_{\rm Dept.}\sim(\mbox{curvatures})+
(\mbox{curvatures})\times(\mbox{fields})+
(\mbox{curvatures})\times(\mbox{fields})^2+\cdots\ .
\label{eightynine}
\ee
Note that one must be somewhat careful when dealing with
derivatives on dependent fields (for example
the constraint $R(Z)^m={}_*R(V)^m$  produces terms $d\dehat z^m$
and $d\dehat v^m$). We count a derivative as field 
number zero so that at each order we cancel all derivative terms.
Note, however, that often one can convert expressions
involving explicit derivatives into expressions 
involving only curvatures and fields. However, 
as we shall soon see, at
the lowest order we are able to remove all terms involving
explicit derivatives. 

We have explicitly
checked the consistency of the expansion described above
by carrying out the analagous calculation in the completely
understood context of conformal supergravity, in order to verify
``unsure'' local supersymmetry at leading order.
 
Let us begin our lowest order (``linear'') analysis.
To compute the variation of the action with respect to $\de$
we need expressions for both $\de_{\rm Group}\calL$ and
$\dehat\calL$. In appendix~\ref{app} we compute expressions 
for both $\de_{\rm Group}\calL$ and $\dehat\calL$ valid to all
orders in our iterative expansion. All terms in 
$\de_{\rm Group}\calL$ (see\rf{vgp}-(\ref{Egp})) are of the form 
(curvature) $\times$ (curvature), since curvatures rotate
homogeneously under the group law, and are of leading order.
The lowest order terms in $\dehat \calL$ (see\rf{dehatS}) 
are those of
the form (curvature)$\times$(vierbein)$\times$ $\dehat$(dept.
field), where, to linear order, the extra transformations
of dependent fields yield curvatures (as in\rf{eightynine} 
above).
In order that the action be invariant at
leading order, terms quadratic in curvatures in $\dehat \calL$
in\rf{dehatS}
must cancel the group transformations of the
action $\de_{\rm Group} \calL$ in~(\ref{vgp})-(\ref{Egp}). 
To lowest order,
the variation of the action is given by
\bea
-\de \calL\!\!\!&=&\!\!\!
-\de_{\rm Group}\calL\!+\!32R(V)^m\de_{\rm
Group}[R(Z)_m\!-\!{}_*R(V)_m]
\!-\!16\sqrt{2}\ \overline{R(S)}\g^5\g^me_m\dehat \phi
\nn\\
&&+32R(K)^me^n\dehat\wt{\o}_{mn}
-128\wt{R}(F)_{mn}e^m\dehat v^n-128R(F)_{mn}e^m\dehat
z^n\nn\\&&\ \ \ 
+O\left([\mbox{curvature}]^2\times (\mbox{field})\right)
\label{linear}
\eea
which is obtained by keeping only the terms with an explicit
vierbein, along with the first two terms in\rf{dehatS}.
Notice, as promised, by virtue of various manipulations made
on the expression for $\dehat \calL$ as given in
appendix~\ref{app}, only  
extra transformations of dependent fields arising from 
constraints without derivatives on dependent fields are needed
(i.e., at linear order, $\dehat v^m$ and $\dehat z^m$ are
determined from the $R(E)^{mn}$ constraints\rf{mc2}-(\ref{mc25}),
$\dehat \o^{mn}$ from $R(P)^m=0$ in\rf{mc1} and $\dehat\phi$ from
$\g^\m R(Q)_{\m\n}=0$ in\rf{mc4}).
Also observe that even though the dilation gauge field $b_\m$ 
may be dependent, at linear level it makes no contribution to
the extra variations of the action.

To verify that the action is invariant at leading order under 
$V_m$ and $Z_m$ symmetries, it only remains to insert the 
explicit
leading order results for the extra transformations of dependent 
fields into $\dehat \calL$ in\rf{linear} and calculate
the group variation of the constraint 
$R(Z)_m-{}_*R(V)_m$ appearing in the first two terms
of\rf{linear} and then compute the sum
$\de_{\rm Group}\calL+\dehat\calL$.
To this end let us give the general, leading order
results for the
extra transformations of the fields $v_\m{}^m$, $z_\m{}^m$,
$\w_\m{}^{mn}$, $b_\m$ and $\phi_\m$ obtained by requiring 
invariance of the maximal set of constraints. 
(The corresponding results
for $f_\m{}^m$ and $F_\m{}^{mn}$ are easily calculated but are
not needed for our linear analysis.) 
\bea
\dehat v_{(\m\n)}&=&-\frac{1}{4}\de_{\rm
Group}\left[R(E)_{\r(\m\n)}{}^\r-\frac{1}{6}g_{\m\n}
R(E)_{\r\s}{}^{\s\r}\right]\\
\dehat z_{(\m\n)}&=&-\frac{1}{4}\de_{\rm
Group}\left[\wt{R}(E)_{\r(\m\n)\r'}g^{\r\r'}-\frac{1}{6}g_{\m\n}
\wt{R}(E)_{\r\s\s'\r'}g^{\r\r'}g^{\s\s'}\right]\\
\dehat z_{[\m\n]}&=&-\frac{1}{4}\a{}_*(\de_{\rm Group}
R(E)_{\r[\m\n]}{}^\r)\label{Anschke}\\
\dehat v_{[\m\n]}&=&-\frac{1}{4}(1-\a)\de_{\rm Group}
R(E)_{\r[\m\n]}{}^\r\label{Katch}\\
\dehat \o^0_{\m mn}&=&-\frac{1}{2}\de_{\rm Group}\left[
R(P)^0_{mn\m}-R(P)^0_{\m mn}+R(P)^0_{\m nm}\right]\\
\dehat \o_m &=& (1-\b) \de_{\rm Group}R(P)_m\label{silk}\\
\dehat b_\m &=& \frac{1}{6}\b \de_{\rm
Group}R(P)_\m\label{pearl}\\
\dehat \phi_\m &=& \frac{1}{2\sqrt{2}}\de_{\rm Group}\left[
\g^\r
R(Q)_{\m\r}+\frac{1}{6}\g_\m\g^{\s\r}R(Q)_{\r\s}\right].
\eea
Here we have denoted the traceless parts of $R(P)_{\m\n m}$ and 
$\o_{\m mn}$ by $R(P)_{\m\n m}^0=R(P)_{\m\n
m}+\frac{2}{3}R(P)_{[\m} e_{\n]m}$ and $\o_{\m mn}^0=
\o_{\m mn}+\frac{2}{3}e_{\m[m}\o_{n]}$, respectively, 
where the traces are denoted by $R(P)_\m=R(P)_{\r\m}{}^\r$ and
$\o_m=\o_{\r m}{}^\r$. Furthermore, since we are working at the
linear level, we can ignore the action of $\de_{\rm Group}$ on
the vierbein and metric\footnote{Of course, one must
nonetheless still resist the temptation to impose constraints
``under'' the $\de_{\rm Group}$ sign. For example, 
$\de_{\rm Group}
\wt{R}(E)_{\m\n\r\s}\equiv
\frac{1}{2}e\e_{\r\s\a\b}\de_{\rm
Group}R(E)_{\m\n}{}^{\a\b}
\neq\de_{\rm Group}{}_*R(E)_{\m\n\r\s}\equiv
\frac{1}{2}e\e_{\m\n\a\b}\de_{\rm
Group}R(E)^{\a\b}{}_{\r\s}$.}, 
because this produces terms that are of
higher order in our expansion. Finally note that we have
introduced two free parameters $\a$ and $\b$ into the results
for the \underline{6} of $v_\m{}^m$ and $z_\m{}^m$ and into
the trace of the spin connection and dilaton, respectively.
Since the constraints only specify the extra transformations of
the combinations $z_{[\m\n]}+{}_*v_{[\m\n]}$ and
$b_\m+\frac{1}{6}\o_\m$ the
results\rf{Anschke},\rf{Katch},\rf{silk} and\rf{pearl}
represent the general solutions for the extra transformations of
these fields. Note that the combinations 
$\a{}v_{[\m\n]}+(1-\a){}_*z_{[\m\n]}$
and $(1-\b)b_\m-\frac{1}{6}\b\o_\m$ are {\it independent}
fields. For example, in conformal supergravity, the choice
$\b=0$ was taken, although this freedom, in any case, cancelled
completely in that model. 
We shall study the dependence of the model on these
parameters in the following calculations.

Applying these results to $V_m$ and $Z_m$ symmetry we find
\bea
\dehat_{V}\o^0_{\m mn}=2R(E)_{mn\m k}\e^k&;&
\dehat_{Z}\o^0_{\m mn}=2\wt{R}(E)_{mn\m k}\e^k\label{one}\\
\dehat_{V}\o_m=0=\dehat_{V}b_\m\hspace{.64cm}&
;&\hspace{.5cm}
\dehat_{Z}\o_m=0=\dehat_{Z}b_\m\\
\dehat_{V}v_\m{}^m=0=\dehat_{V}z_\m{}^m\hspace{.4cm}
&;&\hspace{.4cm}
\dehat_{Z}v_\m{}^m=0=\dehat_{Z}z_\m{}^m\label{twopttwo}\\
\dehat_{V}\phi_\m=\frac{1}{\sqrt{2}}R(Q)_{\m\n}\e^\n
\label{three}&;&
\dehat_{Z}\phi_\m=\frac{1}{\sqrt{2}}\g^5R(Q)_{\m\n}\e^\n,
\label{four}
\eea
Observe that all dependence on the parameters $\a$ and $\b$ has
dropped out.
To obtain\rf{one} we used the cyclicity relation
$R(E)_{[\m\n\r]\s}=0$ in\rf{Ecyc} and 
$\wt{R}(E)_{[\m\n\r]\s}=0$ from\rf{Ecyc1}. 
Reusing these cyclicity
relations and the additional cyclicity
relation $\g_{[\m}R(Q)_{\r\s]}=0$ 
in\rf{cycle2}, it easy to recast 
the results\rf{one}-(\ref{four}) into form notation
\bea
e^n\dehat_{V}\wt{\o}_{mn}&=&-2\wt{R}(E)_{mn}\e^n\label{me}\\
e^n\dehat_{Z}\wt{\o}_{mn}&=&2R(E)_{mn}\e^n\\
e^m\g_m\dehat_{V}\phi&=&-\frac{1}{\sqrt{2}}\g_mR(Q)\e^m\\
e^m\g_m\dehat_{Z}\phi&=&+\frac{1}{\sqrt{2}}\g^5\g_mR(Q)\e^m
.\label{you}
\eea
Let us orchestrate the above formul\ae. Using\rf{vgp},
(\ref{zgp}) and\rf{me}-(\ref{you})
in\rf{linear} we have
\bea
-\de_{V}\calL&=&32[\wt{R}(M)^{mn}R(V)_m-2R(D)R(Z)^n]\e_n\nn\\
            &&+32R(V)^m\de_{V,\rm
Group}[R(Z)_m-{}_*R(V)_m]\label{smythe}\\
-\de_{Z}\calL&=&32[\wt{R}(M)^{mn}R(Z)_m+2R(D)R(V)^n]\e_n\nn\\
            &&+32R(V)^m\de_{Z,\rm Group}[R(Z)_m-{}_*R(V)_m].
\label{prickard}
\eea
Now it only remains to calculate the group transformations of the
constraint\rf{mc3} which, to leading order, yields
\bea
\hspace{.5cm}\de_{V,\rm
Group}[R(Z)_m-{}_*R(V)_m]&=&-\wt{R}(M)_{mn}\e^n+2{}_*R(D)\e_m\nn
\\&&\hspace{.5cm}
\label{alpha}\\
\hspace{.5cm}\de_{Z,\rm
Group}[R(Z)_m-{}_*R(V)_m]&=&R(M)_{mn}\e^n-2R(D)\e_m\nn
\\&&\hspace{.5cm}
\label{beta}
\eea
To obtain these results we have used that, to lowest order,
the following constraints hold
\bea
R(M)^{mn}&=&-{}_*\wt{R}(M)^{mn}\label{him}\\
R(A)&=&{}_*R(D)
\label{her}
\eea
The first may be derived by noting that the $\underline{9}$ and
$\wt{\underline{9}}$ of $R(M)_{\m\n}{}^{mn}$ vanish at
lowest order (although one may verify that at the next to
leading order the quadratric terms 
involving $R(E)^{mn}$, $v^m$, $z^m$ and $E^{mn}$, 
$R(V)^m$, $R(Z)^m$ in\rf{mc5} and\rf{Bianid} 
do contribute to the $\underline{9}$ and $\wt{\underline{9}}$).
Inserting\rf{alpha} and\rf{beta} in\rf{smythe} and\rf{prickard},
respectively, and
using\rf{him} and\rf{mc3}, {\it we see that 
the lowest order contributions
to $\de_{V}\calL$ and $\de_{Z}\calL$ do indeed vanish}.

The invariance of the action under $V_m$ and $Z_m$
symmetries confirms our proposed maximal set of
constraints. Of course, our linear analysis does not probe
the terms in the constraints of higher order in fields.
In the case of the sure constraints however, such
additional terms are severely restricted by the requirement that
the action is invariant under the sure symmetries.
In general, they are also further restricted by requiring
that the constraints themselves are invariant under the sure
symmetries. Further, let us note that the only
(linearized) constraint, which was not needed in the above
analysis is the constraint on the scalar curvature
(\underline{1})
of $R(E)_{\m\n}{}^{mn}$ (i.e. $R(E)_{\m\n}{}^{\n\m}=0$),
although even this constraint is necessary since it follows from
$\g^\m R(Q)_{\m\n}=0$ by varying with respect to the sure
symmetry $S$.
We also observe that the freedom to choose the dependent 
combinations of $z_{[\m\n]}$ and ${}_*v_{[\m\n]}$, and 
$b_\m$ and $\o_\m$ (denoted by the parameters $\a$ and $\b$,
respectively) cancels in the $V_m$ and $Z_m$ variation of the
action at linear order.  
We now turn to the remaining unsure symmetries $Q$ and
$E_{mn}$.

\section{Local Supersymmetry and $E_{mn}$ Symmetry.}
\label{blueline}

These symmetries are of vital importance for the perturbative
consistency of our action, since in a flat 
gravitational and otherwise
trivial background the fields $\psi^\a$ and $E^{mn}$ enter the
quadratic part of the action only in terms 
of their linearized field
strengths $d\psi^\a$ and $dE^{mn}$.  
If the associated gauge
invariances do not survive at the 
interacting level (and they don't),
this kind of perturbation expansion will not lead to invertible
kinetic terms. Conceivably there might 
exist a vacuum solution that
does allow a perturbation expansion. 
However, unless modifications of the supersymmetry and
$E_{mn}$ transformation rules exist under which the action 
is invariant, it is more likely that the theory simply does not
exist as a gauge theory. 
In section~\ref{concl} we examine in more detail some
possibilities for such modifications.

The manipulations required to show that $E_{mn}$ and $Q$ are not
linear order symmetries of the action are 
exactly the same as discussed in detail for $V_m$ and $Z_m$
symmetries,
so let us just give the main results. 

In fact many terms in both $\de_Q\calL$ and $\de_{E}\calL$
do cancel and the only terms remaining are given by
(in form notation)
\bea
-\de_Q\calL\!\!&=&\!\!-16\sqrt{2}\ 
\overline{R(S)}\g^5\g^me_m\dehat_Q\phi
\!-\!
128\wt{R}(F)_{mn}e^m\dehat_Q v^n\!-\!128R(F)_{mn}e^m\dehat_Q z^n
\nn\\&&\label{carl}\\
-\de_E
\calL\!\!&=&\!\!
164R(K)^m[\wt{\e}_{mn}R(V)^n-\e_{mn}R(Z)^n+\frac{1}{2}e^n
\dehat_{E}\wt{\o}_{mn}]
\nn\\&&\hspace{-.6cm}-128\wt{R}(F)_{mn}e^m\dehat_{E} 
v^n-128R(F)_{mn}e^m\dehat_{E} z^n
-16\sqrt{2}\ \overline{R(S)}\g^5\g^me_m\dehat \phi,\nn\\
\label{perkins}
\eea
where the relevant extra transformations of dependent fields 
are given (in explicit index notation) by 
\bea
\hspace{-.5cm}\dehat_Q\phi_\mu&=&
\frac{1}{3\surd 2}[R(D)_{\m\r}+R(A)_{\m\r}\g^5
+\frac{1}{2}(R(V)_\r+R(Z)_\r\g^5)\g_\m]\g^\r\label{sun}\\
\hspace{-.5cm}\dehat_Qv_{\m\n}&=&\frac{1}{8\surd
2}(1-\a)\overline{R(Q)}\e\\
\hspace{-.5cm}\dehat_Qz_{\m\n}&=&\frac{1}{8\surd
2}\a\overline{R(Q)}\g^5\e\\
\hspace{-.5cm}\dehat_{E}\phi_\m&=&
-\frac{1}{4}[\g^\r\g^{mn}R(S)_{\m\r}
+\frac{1}{6}\g_\m\g^{\s\r}\g^{mn}R(S)_{\r\s}]\e_{mn}\\
\hspace{-.5cm}\dehat_{E}\o^0_{\m mn}&=&R(V)_{mn}{}^{r}\e_{r\m}-
                         R(V)_{\m m}{}^{r}\e_{rn}+
                         R(V)_{\m n}{}^{r}\e_{rm}\nn\\&&
         \hspace{-.33cm}+R(Z)_{mn}{}^{\r}{}_*\e_{\r\m}-
                         R(Z)_{\m m}{}^{r}\wt{\e}_{rn}+
                         R(Z)_{\m n}{}^{r}\wt{\e}_{rm}
                        -\frac{2}{3}e_{\m[m}\dehat\o_{n]}\\
\hspace{-.5cm}\dehat_{E}\o_\m&=&2(1-\b)(\wt{\e}_{mn}R(Z)^n+
\e_{mn}R(V)^{n})\\
\hspace{-.5cm}\dehat_{E}v_{[\m\n]}&=& 
\frac{1}{8}(1-\a)[R(M)_{\r\s\m\n}
\e^{\r\s}+8R(D)_{\r[\m}\e_{\n]}{}^\r]\\
\hspace{-.5cm}\dehat_{E}z_{[\m\n]}&=& 
\frac{1}{8}\a
[R(M)_{\r\s\m\n}{}_*\e^{\r\s}
+8R(D)_{\r[\m}{}_*\e_{\n]}{}^\r] \\
\hspace{-.5cm}\dehat_{E}v_{(\m\n)}&=&
\frac{1}{6}g_{\m\n}R(D)_{\r\s}\e^{\r\s}\\
\hspace{-.5cm}\dehat_{E}z_{(\m\n)}&=&
\frac{1}{6}g_{\m\n}R(D)_{\r\s}{}_*\e^{\r\s},
\label{elvis}
\eea
where, as usual, we have denoted the traces
$R(V)_{\n}=R(V)_{\r\n}{}^\r$ and $R(Z)_{\n}=R(Z)_{\r\n}{}^\r$.

In deriving the results\rf{sun}-(\ref{elvis}) we used the
following cyclicity relations which follow directly from 
$R(Z)^m={}_*R(V)^m$,
\bea
R(V)_{\m\r\s}+R(V)_{\r\s\m}+R(V)_{\s\m\r}&=&-e\e_{\m\r\s\tau}
R(Z)^\t\\
R(Z)_{\m\r\s}+R(Z)_{\r\s\m}+R(Z)_{\s\m\r}&=&e\e_{\m\r\s\tau}
R(V)^\t
\eea
It is tedious, but straightforward, to explicitly substitute the
results\rf{sun}-(\ref{elvis}) into\rf{carl} and \rf{perkins}
and verify that the result is indeed non-vanishing. We have 
checked this explicitly.
Let us demonstrate this in a few examples, which, at the same
time, will give us the dependence on the freedoms $\a$ and $\b$.
The curvatures $R(S)_{\m\n}$, $R(K)_{\m\n}{}^m$ and
$R(F)_{\m\n}{}^{mn}$ are all unconstrained, so the coefficients
of each of their Lorentz irreducible parts must vanish
separately (modulo, of course, the maximal set of constraints).

Consider first the $R(F)_{\m\n}{}^{mn}$ terms in $\de_Q \calL$.
Here only the coefficient of the antisymmetric Ricci of
${}_*\wt{R}(F)_{\m\n}{}^{mn}$ is non-zero and is given by
\be
-\de_Q \calL = 8\sqrt{2}\, e\,\overline{R(Q)}_{\m\n}\e
\left[g_{\r\s}{}_*\wt{R}(F)^{\r\m\n\s}\right]+\cdots
\label{notzero}
\ee 
(The dots ``$\cdots$'' denote the remaining term 
proportional to  $R(S)$ 
in $\de_Q\calL$.)
Note, in particular, that the freedom $\a$ to choose which
combination of $z_{[\m\n]}$ and ${}_*v_{[\m\n]}$ is a dependent
field, drops out. We note also that the same freedom $\b$ 
in the $\o_\m$ and $b_\m$ sector also drops out at leading order
(one finds that the extra transformation $\dehat_Q\o_m$ is
identically zero). We observe that since there is no constraint
on $g_{\r\s}{}_*\wt{R}(F)^{\r\m\n\s}$, the expression\rf{notzero}
cannot vanish or cancel with any other terms in $\de_Q\calL$.

Next consider the possibly $\a$ dependent
$R(F)_{\m\n}{}^{mn}$ terms in\rf{perkins}, which are again
proportional to  the antisymmetric Ricci of
${}_*\wt{R}(F)_{\m\n}{}^{mn}$. That $\a$ again drops out
holds, in fact, independently of the particular expression for
$\dehat z_{[\m\n]}$ and $\dehat {}_*v_{[\m\n]}$,
but rather 
since $(1-\a){}_*\dehat z_{[\m\n]}=-\a\dehat v_{[\m\n]}$.
We are left then with the non-zero expression
\be
-\de_{E}\calL=-16 e (R(M)_{\m\n\a\b}
\e^{\a\b}+4R(D)_{\a[\m}\e_{\n]}{}^\a)
\left[g_{\r\s}{}_*\wt{R}(F)^{\r\m\n\s}\right]+\cdots. 
\ee
 
Our final example comprises the terms proportional to the trace
$R(K)^{\r\m}{}_\r$ $\equiv$ $R(K)^\m$ 
in $\de_E\calL$. Here we find that the model actually
retains a dependence on the parameter $\b$. The result is
\bea
-\de_{E}\calL&=&
-\frac{64}{3}(2\b-3)eR(K)^\m[
\e_{\m\n}R(V)^\n+{}_*\e_{\m\n}R(Z)^\n]+\cdots
,\label{Tobias}
\eea
We note that, at leading order, 
this is the only
instance in which the freedom $\a$ or $\b$ does not cancel. 
Furthermore, for the value $\b=3/2$ these terms vanish.
This corresponds to the case in which the combination 
$(b_\m+\frac{1}{2}\o_\m)$ is an independent field. 
Of course, even for this choice of $\b$, the remaining 
terms in $\de_E \calL$ do not vanish.
This concludes our demonstration that the gauge
transformations $Q$ and $E_{mn}$ are not symmetries of the
action at leading order.  

Penultimately,
we make the important remark that invariance at the linear level
is a {\it necessary} but not 
sufficient condition that the action be
invariant under some symmetry. Therefore our analysis shows that
the symmetries $V_m$ and $Z_m$ could possibly be full symmetries
of the action to all orders, but {\it local supersymmetry $Q$ and
$E_{mn}$ are not}. 

Finally, let us conclude this section by remarking that the
sure symmetries act in a very simple way on the
independent fields. For example,
in~\cite{Kaku1} it was shown for the conformal supergravity
model that the conformal boost symmetry
$K_m$ acted only on the dilaton field $b$ as a translation
$b_\mu\rightarrow b_\m-2\e_\m$ whereby it was argued that
the action was independent of $b_\m$. We may perform a similar
analysis here also, but we now find that instead of being
independent of the field $b_\m$, the action depends only on a
certain combination of the independent fields $a_\m$,
$b_\m$ and the remaining independent components of the
$v$ and $z$ fields $z_{[\m\n]}-{}_*v_{[\m\n]}$. 
As shown in the discussion above, after imposing the maximal
set of constraints the remaining independent
fields are $e_\m{}^m$, $E_\m{}^{mn}$, $\psi_\m$, $b_\m$, 
$a_\m$ and the combination $z_{[\m\n]}-{}_*v_{\m\n}$ so that
after solving all constraints 
\be
\calL=\calL(e_\m{}^m,E_\m{}^{mn},\psi_\m,b_\m,a_\m,
            z_{[\m\n]}-{}_*v_{[\m\n]}). 
\ee
Note for simplicity, and indeed without loss of generality that
we have at this point chosen the dilaton $b_\m$ and the
combination $z_{[\m\n]}-{}_*v_{\m\n}$ to be independent 
fields, although as discussed earlier, 
more general combinations 
are possible.
 
Under conformal boosts ($K_m$) the only independent
fields which
transform are $b$, $v^m$ and $z^m$,
\bea
b_\m&\rightarrow&b_\m-2\e_\m\\
z_{[\m\n]}-{}_*v_{[\m\n]}&\rightarrow&
z_{[\m\n]}-{}_*v_{[\m\n]}+2e_{m[\n}\wt{E}_{\m]}{}^{mn}\e_n
-2{}_*[E_{[\m\n]}{}^n\e_n].
\eea
(Since two or more conformal boosts vanish when acting on the
fields $b$, $v^m$ and $z^m$, the transformations given above are
finite gauge transformations rather
than just the infinitesimal transformations.)
Similarly, $F_{mn}$ gauge transformations act only on 
the fields $a$, $b$, $v^m$ and $z^m$ and the finite results are
given by
\bea
a_\m&\rightarrow&a_\m-\wt{E}_\m{}^{mn}\e_{mn}\\
b_\m&\rightarrow&b_\m-E_\m{}^{mn}\e_{mn}\\
z_{[\m\n]}-{}_*v_{[\m\n]}&\rightarrow&
z_{[\m\n]}-{}_*v_{[\m\n]}+4{}_*\e_{\m\n}.
\eea
The action~(\ref{S}) was constructed to be invariant under
conformal boosts ($K_m$) and $F_{mn}$ gauge transformations.
Therefore, one may now construct a $F_{mn}$ gauge 
transformation followed by an appropriate 
conformal boost ($K_m$)
gauge transformation such that 
all dependence on the independent fields $b_\m$ and 
$z_{[\m\n]}-{}_*v_{\m\n}$ appears only in the 
combination $a_\m^\prime=a_\m+\Delta a_\m(b,z-{}_*v)$ so that
$\calL(a_\m,b_\m,z_{[\m\n]}-{}_*v_{\m\n})$ $=$ 
$\calL(a_\m^\prime,0,0)$.
The field equation
equations of the $b_\m$ and $z_{[\m\n]}-{}_*v_{\m\n}$ fields,
have therefore, no more content than that of the $a_\m$ field
equation. We view this $v^m$, $z^m$ and $b$--independent
formulation as a kind of Wess Zumino gauge,
in which we have gauged these fields away. Of course, to
calculate the gauge transformations of the new field
$a_\m^{\prime}$, one must, in the usual fashion, perform
appropriate compensating gauge transformations. Both
formulations are, of course, entirely equivalent, and we found it
more convenient to work in the formulation where the
dependence on the fields 
$v^m$, $z^m$ and $b$ is kept explicit.

\section{Kinematical Analysis and Conclusions.}
\label{concl}

Originally, conformal supergravity was discovered through a
combination of a  
dynamical approach and a kinematical
approach~\cite{Kaku,Kaku1} 
the former being similar to that which we have followed above.
In the dynamical approach, an action 
was constructed, and constraints on curvatures
were found such that the action was invariant under sure and
automatic symmetries. In turn, these constraints endowed the
transformations of dependent fields under ``unsure'' local
supersymmetry with extra pieces and it was verified that 
the modified transformation rules were indeed an invariance of
the action.
The conformal gauge field $f_\m{}^m$ was solved from its own
algebraic field equation, so that its variations did not need to
be taken into account (1.5 order formalism).
However, later it was shown that all 
the results of that model could
also be obtained through an entirely kinematical
approach~\cite{PvN}, in which all constraints were found by
requiring that the gauge algebra closed onto general coordinate
transformations, rather than $P_m$ gauge transformations.

We now apply this kinematical approach to our extended conformal
supergravity model. We find that in order that the gauge algebra
close onto general coordinate transformations, the symmetries 
$Q$ and $E_{mn}$ require extra modifications beyond that implied
by the introduction of constraints on curvatures. This result is
consistent with the findings of our linear analysis in which
$Q$ and $E_{mn}$ were not invariances of the action.

We begin with the important
observation~\cite{MacDowell,Kaku1} 
that general coordinate
transformations
\be
\de_{\rm Gen.\, coord.}h_\m{}^A=\partial_\m\xi^\r h_\r{}^A
+\xi^\r\partial_\r h_\m{}^A,
\ee
and $P_m$ gauge transformations with parameter $\e^m=\xi^\r
e_\r{}^m$ {\it 
differ only by a sum of local gauge invariances of the
theory and a curvature term}
\be
\begin{array}[t]{l}
\de_{\rm Group, P^m} \left(\e^m 
= \xi^\r e_\r{}^m \right)h_\m{}^A
+\xi^\r R_{\r\m}{}^A
\\ 
\hspace{1.8cm}= \de_{\rm Gen.\, coord.} \left( \xi^\r \right)h_\m{}^A
-\de_{\rm Group} \left( \e^{B\neq P_m}
 = \xi^\r
h_\r{}^B\right)
h_\m{}^A.
\end{array}
\label{centraal}
\ee
The last term in\rf{centraal} is a sum of group law gauge
transformations whose field-dependent parameter is given by 
the contraction of the general
coordinate parameter $\xi^\r$ 
with each gauge field $h_\r{}^A$, 
but where $P_m$ gauge transformations are omitted.

Therefore, for the gauge algebra to close, it is sufficient 
that whenever a $P_m$ transformation is produced on the 
right hand side of commutators acting on independent fields, the
combination of any extra terms from the second gauge variation
acting on a possibly dependent field should equal the curvature
term on the left hand side of\rf{centraal}, modulo any
constraints on curvatures.

For example, consider the commutator of two supersymmetry
transformations acting on the independent vierbein field
\be
[\de_Q\left(\e_1\right),\de_Q\left(\e_2\right)]e_\m{}^m=
\de_{P}\left(\frac{1}{2\surd
2}\overline{\e}_2\g^r\e_1\right)e_\m{}^m+
\de_{E}\left(-\frac{1}{2\surd
2}\overline{\e}_2\g^{rs}\e_1\right)e_\m{}^m
\label{smudge}
\ee
Observe that the first supersymmetry transformation
produces an {\it independent} gravitino field $\psi_\m$
so that the result 
above should coincide with the usual group law.
Supposing that $E_{mn}$-gauge transformations were to be
a symmetry of the theory, then the right hand side of 
would also be a symmetry if the constraint
\be
R(P)_{\m\n}{}^m=0
\ee
held (see\rf{centraal}). 
We already found this constraint from the dynamical approach, so
at this point we find a confirmation of our assumptions.
Next we apply the same procedure to the $E_\m{}^{mn}$
gauge field. Again the first supersymmetry transformation
produces the independent gravitino field so that the commutator
of
two supersymmetry transformations yields the group law result
only
\be
[\de_Q\left(\e_1\right),\de_Q\left(\e_2\right)]E_\m{}^{mn}=
\de_{P}\left(\frac{1}{2\surd
2}\overline{\e}_2\g^r\e_1\right)E_\m{}^{mn}+
\de_{E}\left(-\frac{1}{2\surd
2}\overline{\e}_2\g^{rs}\e_1\right)E_\m{}^{mn}
\label{dreck}
\ee
from which we now conclude that the constraint
\be
R(E)_{\m\n}{}^{mn}=0
\label{constraino}
\ee
should hold. However, as discussed in section~\ref{Lorentz},
all of $R(E)_{\m\n}{}^{mn}$ may be solvably 
constrained to vanish {\it except} the traceless \underline{10}.
Therefore, we conclude that in order that the gauge algebra
close, supersymmetry transformations on (some of) the
independent fields should be explicitly
modified. (One may also consider modifying $E_{mn}$
transformations at this point, but since the parameter
$\frac{1}{2\surd 2}\overline{\e}_2\g^r\e_1$
is independent from $-\frac{1}{2\surd2}
\overline{\e}_2\g^{rs}\e_1$ it seems unlikely that a modified
$E_{mn}$ transformation on the right hand side of\rf{dreck}
could produce terms rendering the constraint\rf{constraino}
solvable). 

At this point, it is already clear that the kinematical approach
faces major difficulties, but one can consider also further
commutators acting on independent fields. The remaining 
commutators producing $P_m$ transformations are
$[\de_{V},\de_{E}]$ and $[\de_{Z},\de_{E}]$.  
(It is not necessary to consider the commutator of general
coordinate transformations with dilations or local Lorentz
transformations, since the commutator of a general coordinate
transformation, with a gauge transformation produces the same
gauge transformation, but whose parameter is given by
$\e^A_{[\rm Gen.coord,Group]}=-\xi^\r\d_\r\e^A$). 
Assuming that the constraint
$R(P)_{\m\n}{}^m=0$ holds and that $E_{mn}$ gauge
transformations are a symmetry, acting with commutators 
$[\de_{V},\de_{E}]$ and $[\de_{Z},\de_{E}]$
on the vierbein, one deduces that the extra transformations of 
the dependent $v^m$ and $z^m$ fields under $V_m$ and $Z_m$
symmetries should vanish. Therefore, the kinematical approach,
produces the same results for the extra transformations 
of the spin connection, dilaton and  $v^m$ and $z^m$ fields
under $V_m$ and $Z_m$
symmetries as in the linear analysis of section~\ref{Linear}.
Note, however, that these extra transformations
belong neither to the constraint\rf{constraino} nor the
$R(E)^{mn}$ constraints found in the dynamical approach.
We are therefore forced to conclude that $E_{mn}$ gauge
transformations should also be explicitly modified.

One might consider modifications in the transformation laws of
independent gauge fields proportional to curvatures, so $\de
h\sim \de_{\rm Group}(\e)h + R\e$. For dimensional reasons, such
modifications cannot occur in $\de e^m$ and $\de E^{mn}$, but in
$\de \psi$ one might try a term $\de_Q^\prime \psi\sim
R(E)_{\m\n}{}^{mn}\g_{mn}\g^\n\e_Q$. Since 
only the \underline{10}
in $R(E)_{\m\n}{}^{mn}$ is nonvanishing, this term vanishes,
hence also $\de_Q\psi_\m$ is unmodified. Instead,
one might study $\de_E^\prime v_\m{}^m\sim
R(E)_{\m\n}{}^{mn}(\e_{_{\scriptstyle E}})_n{}^\n$ 
etcetera . However, rather than
searching for modifications of $Q$ and $E_{mn}$ symmetries such
that the gauge algebra closes onto general coordinate
transformations, one suspects that these symmetries should be
dropped altogether, just as was the case for $P_m$ gauge
transformations, and instead there should exist in their place a
generalization of general coordinate transformations to $Q$ and
$E_{mn}$ symmetries with parameters $\xi^\a$ and $\xi^{\m\n}$,
respectively.
One could then study
generalizations of\rf{centraal} to $Q$ and $E_{mn}$ symmetries
in which the new ``general coordinate'' $Q$ and $E_{mn}$
transformations are given by $\de_{{\rm Group}(Q,E_{mn})}$ with
a field dependent parameter $\e\sim\xi h$ 
(which may allow
for a more general mass dimension of the parameter
$\xi$)
plus curvature terms. Perhaps one
can learn more about such a proposal by making a linearized
analysis of the $P_m$ gauge symmetry in the dynamical approach,
since we at least know for certain that the $Osp(1|8)$ model
is general coordinate invariant. We feel that our dynamical
approach has laid the groundwork for such investigations, which
we, however, reserve for further study.

Having shown that the kinematical approach at least produces
results consistent with the findings of the dynamical approach
considered in the text of this paper, let us present our
conclusions and some more speculative remarks.

A simple possibility is that there exists no 
$R^2$ type of supergravity based on $Osp(1|8)$. Let us proceed 
under the {\it assumption} that this is not the case.  
The most obvious way to proceed is then to attempt to combine
the kinematical approach with the solely dynamical approach in
the text in order to find explicit modifications or
generalizations of the
$Q$ and $E_{mn}$ transformation rules, as discussed above,
such that the gauge
algebra closes and invariance of the action is achieved.
The fascinating possibility, that such generalizations of
general coordinate transformations could exist, and can be
probed via a dynamical analysis as given in the text, was a key
motivation for us to study this model in detail, even though,
{\it a priori}, from a kinematical standpoint, we suspected
that new features would arise in the $Q$ and $E_{mn}$
symmetries.  

In the limit that the symmetries become rigid and $e_\m{}^m$
becomes a flat space delta function $\de_\m{}^m$, one would
expect that one should add orbital parts to the transformation
rules which are a consequence of the transformation of
coordinates.
The most general set of bosonic rigid symmetries leaving the line
element $(dx^m)^2$ invariant is known to be the conformal
group. Clearly, $E$ symmetry does not act on the coordinates
$x^m$. An alternative which we have not pursued in this article
at all, is to consider a space with bosonic coordinates $x^m$
and $y^{[mn]}$ ( or a superspace with $x^m$, $\theta^\a$ and
$y^{[mn]}$). One might then study  
dimensional reduction of a higher
dimensional model. For example, in order to construct 
rigid representations of the superalgebra 
$Osp(1|8)$, one can first 
consider the set of non-negative dilaton weight generators
$\{P_m,E_{mn},Q;M_{mn},D,A,V_m,Z_m\}$ and consider a coset
based on the reductive split where the positive dilaton
weight generators are coset generators and the zero weight
generators are subgroup generators. (Representations of the full
$Osp(1|8)$ superalgebra should 
then be calculated using the theory
of induced representations).
One must consider then a superspace with coordinates
$(x^m,y^{[mn]},\theta^\a)$ where $x^m$ are the four
spacetime coordinates, $\theta^\a$ ($\a=1,4$)
are the usual anti-commuting
coordinates and the $y^{[mn]}$ are six auxiliary coordinates. 
Notice that one does not have a ten dimensional Lorentz
group, but rather, 
the auxiliary coordinates $y^{[mn]}$ transform as a rank two
antisymmetric tensor under four dimensional 
Lorentz transformations. 
In two component notation the anti-commutator of two $Q$
transformations reads
\bea
\{Q_A,Q_B\}&=&E_{(AB)}\label{a1}\\
\{Q_A,Q_{\dot{B}}\}&=&P_{A\dot{B}}\\
\{Q_{\dot{A}},Q_{\dot{B}} \}&=&E_{(\dot{A}\dot{B})},\label{a2}
\eea
where $A,B,\dot{A},\dot{B}=1,2$ and we have made the usual
decomposition of the rank two antisymmetric tensor $E_{mn}$
into its  self-dual and anti-self-dual parts.
The algebra\rf{a1}-(\ref{a2}) represents a natural extension of
the usual supersymmetry algebra but further 
possible constraints and/or dimensional reduction schemes are
needed to
to control dependence on the six auxiliary coordinates
$y^{[mn]}$ such that one regains new representations of
$Osp(1|8)$ in four dimensions. 

\section{Acknowledgements}

We heartily thank Warren Siegel, Kostas Skenderis,
Koenraad Schalm and Michail A. Vasiliev 
for useful discussions.

\vfill

\appendix

\section{Group Law and Extra Transformations of the Action under
Unsure Symmetries.}
\label{app}

In this appendix we derive complete
expressions for both $\de_{\rm Group}\calL$ and
$\dehat\calL$ valid to all orders in fields for some 
(unsure) symmetry $\de$. 
The former is trivial to compute since curvatures
just rotate homogeneously according to the group law
(see\rf{hiatus})
and the results for all  unsure symmetries are
\bea
\hspace{0cm}\!\!\!-\de_{V, \rm
Group}{\cal
L}&\!\!=\!\!&16\begin{array}[t]{l}[2\wt{R}(M)^{mn}R(V)_m-
4R(D)R(Z)^n\\
                +4\wt{R}(E)^{mn}R(K)_m
                 -\overline{R(Q)}\g^5\g^nR(S)]\e_n
\end{array}\label{vgp}\\
\hspace{0cm}\!\!\!-\de_{Z, \rm Group}{\cal L}
&\!\!=\!\!&16\begin{array}[t]{l}[2\wt{R}(M)^{mn}R(Z)_m+
4R(D)R(V)^n\\
                -4R(E)^{mn}R(K)_m
                 +\overline{R(Q)}\g^nR(S)]\e_n
\end{array}\label{zgp}\\
\hspace{0cm}\!\!\!-\de_{Q, \rm Group}{\cal L}
&\!=\!\!&8\!\begin{array}[t]{l}
[\overline{R(S)}\{\!\g^5\g_mR(V)^m\!
+\!\g_mR(Z)^m\}
                +\sqrt{2}\ 
\overline{R(Q)}\g^5\g_mR(K)^m\!]\e\end{array}\nn\\&&\\
\hspace{0cm}\!\!\!-\de_{P, \rm
Group}{\cal
L}&\!\!=\!\!&-32\begin{array}[t]{l}[
\wt{R}(M)^{mn}R(K)_m+2R(A)R(K)^n
                +2\wt{R}(F)^{mn}R(V)_m\\+2R(F)^{mn}R(Z)_m
 +4\sqrt{2}\ \overline{R(S)}\g^5\g^nR(S)]\e_n
\end{array}\nn\\&&\\
\hspace{0cm}\!\!\!
-\de_{E, \rm Group}{\cal L}&\!=\!\!&0.\label{Egp}
\eea
For completeness we have included the result for
$P_m$ gauge transformations, even though the action, being
general coordinate invariant, will not, in general,
be $P_m$ invariant. It is surprising that the action is invariant
under $E_{mn}$ group transformations but note that 
we still have to compute the effect of the extra 
$E_{mn}$ transformations. 

Let us now compute the variation of the action under
extra transformations $\dehat$. Under the most general
variations of all fields one finds
\vspace{-.1cm}
\bea
-\widehat{\de}{\cal L}&=&
32\begin{array}{l}
 [v^mR(V)^n+z^mR(Z)^n-e^mR(K)^n]\widehat{\de}\wt{\o}_{mn}
\end{array}\nonumber\\
&&\hspace{-.4cm}+32\begin{array}[t]{l}[\wt{R}(M)^{mn}v_m-
                    2\wt{R}(F)^{mn}e_m+2\wt{R}(E)^{mn}f_m\\
  -2R(D)z^n-\frac{1}{4}\psibar
\g^5\g^nR(S)-\frac{1}{4}\overline{R(Q)}\g^5
\g^n\phi]\widehat{\de}v_n\end{array}
\nonumber\\
&&\hspace{-.4cm}+32\begin{array}[t]{l}[\wt{R}(M)^{mn}z_m
              -2R(F)^{mn}e_m-2R(E)^{mn}f_m\\
    +2R(D)v^n+\frac{1}{4}\psibar\g^nR(S)
    +\frac{1}{4}\overline{R(Q)}\g^n\phi]\widehat{\de}z_n
    \end{array}
\nonumber\\
&&\hspace{-.4cm}+16\begin{array}[t]{l}[\overline{R(Q)}
(\g^5v^m\g_m-z^m\g_m)-\frac{1}{2}\psibar(\g^5
R(V)^m\g_m-R(Z)^m\g_m)\\
-\sqrt{2}\ \overline{R(S)}\g^5e^m\g_m]\widehat{\de}\phi
\end{array}
\nonumber\\
&&\hspace{-.4cm}+32\begin{array}[t]{l}[\wt{R}(M)^{mn}e_n-
          2\wt{R}(E)^{mn}v_n+2R(E)^{mn}z_n\\
    +2R(A)e^m-\frac{1}{2\sqrt{2}}\ 
     \overline{\psi}\g^5\g^mR(Q)]
\widehat{\de}f_m\end{array}
\nonumber\\&&\hspace{-.4cm}
+64\begin{array}{l}[R(V)^me^n\widehat{\de}\wt{F}_{mn}
+R(Z)^me^n\widehat{\de}F_{mn}]
\end{array}\nonumber\\
&&\hspace{-.4cm}-32\begin{array}{l}
[R(V)^mz_m-R(Z)^mv_m]\dehat b+64R(K)^me_m\dehat a
\end{array}\nonumber\\
&&\hspace{-.4cm}-8\begin{array}[t]{l}
    [\overline{\phi}(R(V)^m\g^5\g_m+R(Z)^m\g_m)
    -2\overline{R(S)}(v^m\g^5\g_m+z^m\g_m)\\
    +\sqrt{2}\ \overline{\psi}R(K)^m\g^5\g_m
    -2\sqrt{2}\ \overline{R(Q)}f^m\g^5\g_m]\dehat \psi
\end{array}\nonumber\\
&&\hspace{-.4cm}
-64\begin{array}[t]{l}
    [(R(K)^mv^n-f^mR(V)^n)\dehat \wt{E}_{mn}-
    (R(K)^mz^n-f^mR(Z)^n)\dehat E_{mn}]
   \end{array}\nonumber\\
&&\hspace{-.4cm}
+32\begin{array}[t]{l}
   [\wt{R}(M)^{mn}f_n-2R(A)f^m+2\wt{R}(F)^{mn}v_n\\
   +2R(F)^{mn}z_n+
   \frac{1}{2\sqrt{2}}\overline{R(S)}\g^5\g^m\phi
   ]\dehat e_m.
\end{array}\label{dehat}
\eea
Note that the independent fields, of course, get no extra
transformations so that $0=\dehat a=\dehat\psi=\dehat
e^m=\dehat E^{mn}$ but we have included these
results here for completeness. The above formula was derived
simply by varying all fields $h^A\rightarrow h^A+\dehat h^A$
as they appear in the explicit results for the
curvatures~(\ref{curve})-(\ref{curvF}). However whenever
terms $d\dehat h^A$ occurred, 
we integrated by parts and then used
the Bianchi identity in the form~(\ref{Bia}) to convert 
the exterior derivative on curvatures into a sum of terms of the
form (field)$\times$(curvature). One finds then many
cancellations and the result for
$\dehat \calL$ has the form above involving only terms of
the form (curvature)$\times$(field)$\times\dehat$(dept. field).

The result~(\ref{dehat}) can be brought into a more
useful form by the following manipulations. First the
coefficients of $\dehat f_m$ almost comprise the
constraint~(\ref{feqn}) so that these terms may be rewritten as
$-64R(V)_m\wt{E}^{mn}\dehat 
f_n+64R(Z)_mE^{mn}\dehat f_n$. However,
if we now use the constraint $R(Z)^m={}_*R(V)^m$ to 
convert all $R(Z)^m$ curvatures to $R(V)^m$ curvatures, then
the set of terms depending on $R(V)^m$ is given by
\be
-32R(V)^m\left\{\begin{array}{l}
2\dehat \wt{F}_{mn}e^n+2\wt{E}_{mn}\dehat f^n
+\frac{1}{4}\psibar
\g^5\g_m\dehat\phi-\dehat \wt{\o}_{mn}v^n+z_m\dehat b\\
-{}_{{}_{\textstyle *}}\!\!\left[-2
\dehat F_{mn}e^n+2E_{mn}\dehat f^n
+\frac{1}{4}\psibar
\g_m\dehat\phi+\dehat \wt{\o}_{mn}z^n+v_m\dehat b\right]
\end{array}\right\}.
\label{Tom}
\ee
Now, notice that the variation of the
constraint $R(Z)^m={}_*R(V)^m$
in~(\ref{c3}) yields 
a similar set of terms to those appearing in\rf{Tom}
\bea
\!\!\!0&=&\de[R(Z)_m-{}_*R(V)_m]\nonumber\\
\!\!\!\!\!\!\!&=&\de_{\rm
Group}[R(Z)_m-{}_*R(V)_m]\nonumber\\
&&\hspace{-.1cm}\!\!\!\!+\left\{\!\!\begin{array}{l}
d\dehat z_m+\dehat (\o_{mn}z^n)-2\dehat v_m a+
2\dehat \wt{F}_{mn}e^n+2\wt{E}_{mn}\dehat f^n
+\frac{1}{4}\psibar
\g^5\g_m\dehat\phi\\
-{}_{{}_{\textstyle *}}\!\!\left[d\dehat v_m+\dehat
(\o_{mn}v^n)+2\dehat z_m a-2
\dehat F_{mn}e^n+2E_{mn}\dehat f^n
+\frac{1}{4}\psibar
\g_m\dehat\phi\right]
\end{array}\!\!\right\}\! .\nn\\&&
\label{159}
\eea  
Therefore, we may rewrite~(\ref{dehat}) as 
\bea
-\dehat \calL &=&32R(V)^m\{\de_{\rm
Group}[R(Z)_m-{}_*R(V)_m]\nn\\&&\hspace{1.82cm}
+\dehat\wt{\o}_{mn}v^n+{}_*(\dehat\wt{\o}_{mn}z^n)
-z_m\dehat b+{}_*(v_m\dehat b)\}
\nn\\&&
-128\wt{R}(F)_{mn}e^m\dehat v^n-128R(F)_{mn}e^m\dehat z^n\nn\\&&
-64R(K)^m[E_{mn}\dehat z^n-\wt{E}_{mn}\dehat v^n
-\frac{1}{2}e^n\dehat\wt{\o}_{mn}]\nn\\
&&+64R(D)[\dehat v^mz_m+v^m\dehat z_m]
+64R(A)[z^m\dehat z_m+v^m\dehat v_m]\nn\\
&&+32R(M)^{mn}[-\dehat v_mz_n+v_m\dehat z_n
+\frac{1}{2}\e_{mnpq}(v^p\dehat v^q+z^p\dehat z^q)]\nn\\
&&+16\overline{R(Q)}[\g^5\g^mv_m-\g^mz_m]\dehat \phi
-16\sqrt{2}\ \overline{R(S)}\g^5\g^me_m\dehat \phi\nn\\
&&+16\psibar[\g^m\dehat z_m-\g^5\g^m\dehat v_m]R(S),
\label{dehatS}
\eea
where we have again integrated by parts in the terms
involving $d\dehat v^m$ and $d\dehat z^m$ 
introduced in\rf{159} and used the Bianchi 
identity. Furthermore we observe that the result~(\ref{dehatS})
for $\dehat \calL$
provides a possible avenue to avoid the problem of coupled
expressions for the extra transformations of dependent fields.
Indeed, if one were able to further rewrite~(\ref{dehatS})
in terms only of the combinations of fields and extra
transformations of dependent fields that appear in variations
of the
constraints, one could then express $\dehat\calL$ in terms of 
readily calculable ``group'' transformations of curvatures.
We have investigated this possibility further, but have found no
obvious way in which this can be done.

\end{document}